\documentclass[english,gc, manuscript]{copernicus}
\usepackage[T1]{fontenc}
\usepackage{array}
\usepackage{multirow}
\usepackage{graphicx}
\usepackage{rotfloat}
\usepackage[authoryear]{natbib}

\makeatletter

\providecommand{\tabularnewline}{\\}

\usepackage{colortbl}
\DeclareRobustCommand{\disambiguate}[3]{#2~#3}

\makeatother

\usepackage{babel}
\begin{document}
\title{Evaluating participants' experience of extended interaction with cutting-edge
physics research through the PRiSE `research in schools' programme}
\author[1,{*}]{Martin O. Archer}
\author[2,3]{Jennifer DeWitt}
\author[4]{Charlotte Thorley}
\author[5]{Olivia Keenan }
\affil[1]{School of Physics and Astronomy, Queen Mary University of London,
London, UK}
\affil[2]{Institute of Education, University College London, London, UK}
\affil[3]{Independent Research and Evaluation Consultant, UK}
\affil[4]{Public Engagement and Involvement Consultant, UK}
\affil[5]{South East Physics Network, London, UK}
\affil[{*}]{now at: Space and Atmospheric Physics, Department of Physics, Imperial
College London, London, UK}
\correspondence{Martin O. Archer\\
(martin@martinarcher.co.uk)}
\runningtitle{Physics Research in School Environments}
\runningauthor{Archer et al.}
\maketitle
\nolinenumbers
\begin{abstract}
Physics in schools is distinctly different from, and struggles to
capture the excitement of, university research-level work. Initiatives
where students engage in independent research linked to cutting-edge
physics within their school over several months might help mitigate
this, potentially facilitating the uptake of science in higher education.
However, how such initiatives are best supported remains unclear and
understudied. This paper evaluates a provision framework, `Physics
Research in School Environments' (PRiSE), using survey data from participating
14--18 year-old students and their teachers to understand their experience
of the programme. The results show that PRiSE appears to provide much
more positive experiences than typical university outreach initiatives
due to the nature of the opportunities afforded over several months,
which schools would not be able to provide without external input.
The intensive support offered is deemed necessary, with all elements
appearing equally important. Based on additional feedback from independent
researchers and engagement professionals, we also suggest the framework
could be adopted at other institutions and applied to their own areas
of scientific research, something which has already started to occur.
\end{abstract}

\introduction{}

Research, policy, and practice all agree that participation in Science,
Technology, Engineering and Mathematics (STEM) needs to be increased
and widened \citep[e.g.][]{case14}, with these issues being particularly
acute for the subject of physics \citep[e.g.][]{murphy06,iop14}.
Physics as a field has become within society strongly aligned with
intelligence/cleverness, masculinity and whiteness, all of which can
dissuade school students (even those highly-enthusiastic about the
subject) from pursuing it further and thereby showing inequitable
effects on those from under-represented backgrounds \citep{archer20}.
Some of these issues arise from practices in school-level physics
education. Debarring and gatekeeping of physics based on attainment
(disproportionately so compared to other subjects) simply feeds the
alignment of physics with cleverness and can make even high-attaining
students' confidence in the subject precarious. Teachers and the school
environment often (even unconsciously) reinforce stereotypes about
physics and physicists that are patterned by biases. Curriculum practices
in physics often teach oversimplifications at younger ages which are
later completely reconceptualised without being presented as refinements
to a model, making students perceive the simpler versions as ``lies''.
Furthermore the general deferment of ``interesting'' physics in
the curriculum produces a disconnect between ``school physics''
and ``real physics'', i.e. the cutting-edge research undertaken
by professional physicists, making continued participation in physics
education something of a ``test of endurance''. These concerns are
further reflected in results from national surveys. While $20\%$
of 16--18 year-old physics school students in the UK aspire towards
a physics degree and $80\%$ aspire towards STEM more broadly \citep{wellcome17},
only $9.7\%$ and $59.3\%$ actually go on to study either physics
or STEM respectively \citep{iop12}. These constitute odds ratios
for aspirations vs. destinations of 2.3 and 2.7, both of which are
considerable. All of these issues raised cultivate and contribute
to reproducing inequitable, and low overall, patterns of participation
within physics \citep{archer20}.

\citet{davenport20} suggest that for STEM subjects in general an
intervention approach that sustains and supports science identity
is most appropriate for students in late secondary/high school education
in the context of their educational journey. However, in the case
of physics specifically, \citet{archer20} comment that existing interventions
based on simply enthusing, inspiring and informing students about
physics will not significantly change uptake or diversity in post-compulsory
physics. While they advocate for widespread changes in science education
policy and practice, both at school and university levels, they note
that if interventions are also used they need to fundamentally address
the problematic processes and practices present within both physics
teaching and physics as a field generally.

The stark differences between school, university, and professional
science practices have long been noted --- while research is one
of the main activities of professional scientists, it is quite removed
from how science is taught in schools with some arguing science education
is not ``authentic'' in this respect \citep[e.g.][]{hodson,braund07}.
Indeed, \citet{yeoman17} report that school students are largely
unaware of what research actually is, finding a disconnection between
\textquoteleft research as information gathering\textquoteright{}
and the \textquoteleft research question\textquoteright , and in general
have little opportunity to set their own research questions within
their school environment. Independent research projects, which provide
extended opportunities for students to lead and tackle open-ended
scientific investigations (not simply literature reviews or essays,
e.g. \citealp{conner09,corlu14}), may be one way of exposing students
to ``real science''. These align with established international
pedagogical initiatives such as `inquiry-based science\textquoteright{}
\citep[e.g.][]{minner10}, `problem-based learning\textquoteright{}
\citep[e.g.][]{gallagher95} and `authentic science\textquoteright{}
\citep[e.g.][]{braund07}. A survey of such projects across twelve
countries (Australia, Ireland, Israel, Netherlands, New Zealand, Qatar,
Singapore, Spain, Taiwan, Turkey, UK, and USA), however, found them
to be rare globally and only sometimes supported by mentors from university/industry
\citep{bennett16,bennett18}. This review found considerable variability
in the nature of independent research projects such as their focus,
delivery/provision models, external support, and funding/costs. It
was noted that such programmes place demands on time and money beyond
standard provisions for all stakeholders, on the skills required by
teachers and other adults involved, and on the supporting infrastructure.
For successful projects \citet{dunlop19} recommend that students
should be given the freedom to devise a research question, have ownership
over their own data analysis and decision-making, and be given access
to experts in their project work. Broadly there are two distinct formats
of independent research projects:
\begin{itemize}
\item Those associated with dedicated out-of-school events of only a few
weeks' duration such as internships/apprenticeships (e.g. `Nuffield
Research Placements' in the UK, \citealp{nuffield19}; `Raising Interest
in Science and Engineering' in the USA, \citealp{stanford}), summer
schools/camps (e.g. the `International Astronomical Youth Camp' run
across Europe and parts of Africa, \citealp{dalgleish}), or science
competitions/fairs (particularly prevalent in the USA, e.g. \citealp{yasar03}).
\item Those undertaken within school itself over the course of several months
to a year, either in class or supplemented with time in after-school
clubs (e.g. an after-school mechatronics project in Taiwan, \citealp{hong13};
class-based biology project in Singapore, \citealp{chin10}; or various
`CarboSchools' climate change projects between research institutes
and schools across 7 European countries, \citealp{dijkstra11}).
\end{itemize}
We do not discuss the former here as they are necessarily limited
in reach, catering only to heavily bought-in individuals (i.e. typically
1--3 students; \citealp{paull17}) from any given school. Most independent
research projects based within schools are not linked to current cutting-edge
and novel scientific research topics or questions. However, relatively
recently so-called `research in schools' projects have emerged, which
do provide students with experiences of genuine contemporary STEM
research within their own school environment over several months.
While several citizen science projects have also run within schools
and aim to help participants learn about current science and to experience
the scientific research process, these are typically secondary aims
since they primarily concern a single (or small number) of well-defined
science questions which will be assisted through developed citizen
science protocols \citep{bonney09,bonney16,shah16}. This contrasts
with independent research projects, and thus also `research in schools',
where positively affecting the participants is the primary concern
and the projects are necessarily open-ended. Nonetheless, the different
approaches can have some overlap and indeed some projects denoted
as citizen science, such as the `curriculum-based' projects based
in the USA described by \citet{bonney16}, might perhaps be better
framed as `research in schools'. `Research in schools' programmes
appear at present to be most prevalent in the UK and we are aware
of three featuring projects in the physical sciences (outside of that
at Queen Mary University of London, QMUL, which forms the subject
of this paper).

HiSPARC (High School Project on Astrophysics Research with Cosmics)
is a scintillator-photomultiplier cosmic ray detector project originating
in the Netherlands at Radboud University in 2004, which has subsequently
been adopted by other Dutch (Amsterdam, Eindhoven, Groningen, Leiden,
Nikhef, Twente, Utrecht) as well as UK (Bath, Bristol, Birmingham,
Sussex) and Danish (Aarhus) universities \citep{colle07,vandam20,hisparc}.
Many of these universities operate a tiered membership scheme for
schools: `Gold' enables schools to buy their own detector (\textsterling5,500
in the UK); `Silver' is a detector rental scheme (\textsterling300
p.a. plus an installation fee) with the contract specifying if they
do not participate the detector will be collected with an additional
fee; and `Bronze' membership (\textsterling200 p.a.) gives schools
access to HiSPARC data but not their own detector. Schools signing
up for `Silver' or `Bronze' membership are contractually obliged to
generate funding to upgrade to `Gold'. While the `Gold' membership
fee covers the costs of the detector and installation, the other memberships
are to ensure that schools make a commitment to working with the university
\citep[J. Velthuis and M. Pavlidou, personal communication, 2016;][]{hestemswr19}.
It is not possible to compare how these schools go about project work
and how much support they are given by participating universities,
which may vary by institution, as at the time of writing HiSPARC has
not published any reviews of their processes or evaluation.

IRIS (Institute for Research in Schools) is a UK charity formally
launched in March 2016 \citep{iris20}, building on the previous CERN@School
project conceived in 2007 \citep{whyntie16,parker19}. While IRIS's
projects cover all the sciences, current physics projects include
the aforementioned CERN@School, Higgs Hunters \citep{barr18}, LUCID
\citep{furnell19,hatfield19}, and Webb Cosmic Mining (in preparation
for the James Webb Space Telescope). They have rapidly expanded across
the UK since formation, having worked in some capacity with over 230
schools as of 2020. Publications have provided technical details of
their projects and case studies of some students' successes within
them, including a few examples of resulting peer-reviewed scientific
work, however the exact provision/delivery model implemented and precisely
how project work is supported is not fully explored in the available
literature. IRIS aims to develop `teacher scientists', teachers that
identify as both science teachers and research-active scientists \citep{rushton19},
which suggests a teacher-driven model. While some researchers/academics
have designed or consulted on some IRIS projects, they appear in general
to have little involvement supporting students or teachers (O. Moore,
personal communication, 2020) with IRIS itself being the main point
of contact for schools. With a recent change of staff at IRIS in late
2019 has come a reformulation of how they classify their projects.
`Seed' projects are for new schools, are the most straightforward,
and receive the most support from IRIS though it is not clear in what
form that takes. `Sprout' projects are more advanced seeing students
carrying out more complex activity to assist scientists with their
research questions, though how this collaboration operates is not
specified. `Grow' projects are where students have proposed their
own research questions, either independently or using IRIS resources,
with IRIS merely providing advice in producing posters, talks, or
papers as well as opportunities to present.

ORBYTS (Original Research By Young Twinkle Scientists), based at University
College London, was piloted from January 2016 and is nominally based
around the Twinkle mission, though has expanded into other research
areas since \citep{sousasilva17,orbyts}. A select group of students
(with an imposed limit of 4--6) from each school undertakes fortnightly
meetings with early career researchers (either PhD students or post-docs)
throughout their project aiming to achieve, where possible, publishable
scientific results \citep{mckemmish17,chubb18,holdship19}. Teachers,
while present, are not typically actively involved in these sessions
and students tend to do little independent work outside of the sessions
(W. Dunn, personal communication, 2018). The content of the projects
change each year to align with the researchers' current focus, with
them typically working with only one school per year each. PhD researchers
are paid for their (preparation, travel, and session) time with funds
from independent schools, who pay not only for their school but for
enabling an additional school from a lower socio-economic background
to take part.

It is clear that there is currently a lack of published details and
evaluation on provision within the emerging area of `research in schools'.
This paper therefore explores these aspects applied specifically to
the `research in schools' programme of QMUL's School of Physics \&
Astronomy. This was piloted between 2014--2016, as detailed in \citet{archer_report17},
and is now known as `Physics Research in School Environments' \citep{prise}.
Section~\ref{sec:PRiSE-framework} introduces PRiSE's framework,
which is then evaluated in section~\ref{sec:Feedback} in terms of
participating students' and teachers' experience. We also briefly
discuss how the PRiSE approach has been received by the university
sector in section~\ref{sec:Future}.

\section{PRiSE framework\label{sec:PRiSE-framework}}

`Physics Research in School Environments' (PRiSE) is a collection
of physics-based `research in schools' projects (see Table~\ref{tab:projects})
brought together under a coherent provision framework. The programme
aims to equip 14--18 year-old school students (particularly those
from disadvantaged backgrounds) with the ability, confidence, and
skills to increase/sustain their aspirations towards physics or more
broadly STEM, ultimately enabling them to realise these at higher
education and thus contributing to increased uptake and diversity
of physics, and to some extent STEM \citep[cf.][]{archer20}. Through
working with teachers, PRiSE also aims to develop their professional
practice and build long-term university--school relationships that
raise the profile of science and mitigate biases/stereotypes associated
with physics within these schools, generally making them environments
which nurture and enhance all students' science capital \citep[cf.][]{iop14}.
Our rationale for these particular outcomes is left to the supplementary
material. This section summarises the PRiSE framework, discussing
the ethos of the programme, the roles played by the schools and university,
as well as outlining the various activity stages, interventions, and
resources which it consists of. In-depth practical details aimed at
practitioners looking to replicate the framework are given in the
supplementary material.

\begin{table*}
\begin{footnotesize}%
\begin{tabular}{>{\raggedright}p{0.25\columnwidth}lll>{\raggedright}p{0.33\columnwidth}}
\textbf{Project} & \textbf{Abbreviation} & \textbf{Years} & \textbf{Field} & \textbf{Description}\tabularnewline
\hline 
\noalign{\vskip6pt}
Scintillator Cosmic Ray Experiments into Atmospheric Muons & SCREAM & 2014--2020 & Cosmic Rays & Scintillator -- Photomuliplier Tube detector usage\tabularnewline
\noalign{\vskip6pt}
Magnetospheric Undulations Sonified Incorporating Citizen Scientists & MUSICS & 2015--2020 & Magnetospheric Physics & Listening to ultra-low frequency waves and analysing in audio software\tabularnewline
\noalign{\vskip6pt}
Planet Hunting with Python & PHwP & 2016--2020 & Exoplanetary Transits & Learning computer programming, applying this to NASA Kepler and TESS
data\tabularnewline
\noalign{\vskip6pt}
ATLAS Open Data & ATLAS & 2017--2020 & Particle Physics & Interacting through online tool with LHC statistical data on particle
collisions\tabularnewline
\end{tabular}\end{footnotesize}

\caption{A summary of the existing PRiSE projects at QMUL.\label{tab:projects}}
\end{table*}

\subsection{Approach}

PRiSE takes the `research in schools' approach to schools engagement,
whereby students are given the opportunity to lead and tackle open-ended
scientific investigations in areas of current research. Therefore,
the PRiSE projects were developed to transform current scientific
research methods, making them accessible and pertinent to school students
so that they could experience, explore, and undertake scientific research
themselves.

\subsubsection{Ethical considerations}

Since the programme intends to influence school students and teachers
a number of ethical considerations have been taken into account, following
the \citet{bera18} guidance for educational research, with regard
to safeguarding and to ensure that no harm results. Firstly, to ensure
equality of access to the programme we do not charge schools to be
involved \citep[cf.][]{harrison10,jardinewright12} and try to provide
them with all the physical resources they need for their project,
thereby removing potential barriers to entry for less resourced schools.
Our targeting takes into account several school-level metrics (type
of school, students on free school meals, indices of multiple deprivation,
gender balance etc.) to ensure diversity. We aim for the programme
to be equitable with all schools being offered the same interventions/opportunities,
taking into account and being flexible to their specific needs where
necessary. We work with as many schools as we have capacity to do
so each year and do not withold interventions from any students for
the purpose of having control groups. The projects are optional and
presented as an opportunity that students can take advantage of which
will be supported by their teacher and the university, therefore students
are not pressured into being involved. Students and schools can drop
out at any point within the programme with no penalty, which does
occur in a minority of schools.

\subsubsection{Role of teachers}

The involvement of teachers at all stages is of paramount importance
in terms of delivery and safeguarding. They have helped shape the
design of the programme, inform how we update it each year, and serve
as our liaison to schools and the students involved. It is the teachers
that decide who projects are offered to within their school, with
us simply advising that the projects should be suitable for all A-Level
(16--18 year-old) students as well as high-ability GCSE (14--16
year-old) students (further contextual information on the UK education
system is given in Appendix~\ref{sec:UK-schools}). These recommendations
were made based on the basic background knowledge required to meaningfully
engage with the research. Invariably teachers choose to involve older
age groups, with $79\pm1\%$ of PRiSE students being aged 16--18
(and so far only one student below our recommended ages has been involved,
being 13--14). Unfortunately, we do not have any specific information
on exactly how teachers go about selecting students. However, the
average number of students per school each year is around 12, which
compared to the national average class size in A-Level physics of
16 \citep{iop15} indicates teachers involve a significant majority
(or in many cases the entirety) of their cohorts in PRiSE. We allow
teachers to determine how best to integrate the projects within their
school, though provide advice on this. We also aim, through our resources
and communications, to equip teachers to manage the day-to-day aspects
of the projects without overly burdening them --- their role is chiefly
one of encouraging their students to persist, providing what advice
they can, and then communicating with the university.

Teachers' involvement at all stages also presents opportunities to
them for continuing professional development, helping them nurture
and cultivate STEM aspirations among students throughout their school
(i.e. not just PRiSE students). This is implemented informally and
integrated within the programme in the form of both bespoke resources
and ongoing dialogues between teachers and researchers. These are
aimed to enhance teachers' knowledge about the underlying science
and how they link to curriculum-based topics where appropriate, their
skills and confidence surrounding current research topics and methods,
and their pedagogy in mentoring independent project work.

\subsubsection{Role of the university}

It was recognised that teachers in general likely will not have the
skills or experience in research to manage projects without expert
assistance \citep{shah16,bennett16,bennett18}. Therefore, PRiSE was
designed to be supported by active researchers equipped with the necessary
expertise to draw upon in offering bespoke, tailored guidance to the
students and teachers. Well-defined roles within the university have
been established for each of the PRiSE projects to provide this support:
\begin{itemize}
\item \textbf{Outreach Officer: }Manages the entire programme including
university--school relationships, communications, intervention/event
co-ordination, programme finances and evaluation.
\item \textbf{Project Lead:} Visible figurehead for the project to schools,
typically an academic member of staff.
\item \textbf{Researcher:} Providing advice and guidance to students and
teachers throughout the programme. Can be delegated to or shared with
early career researchers.
\end{itemize}
Since the primary focus of PRiSE is (unlike typical citizen science)
on the participants rather than the research, researchers should not
consider students contributing to novel research as their rationale
for being involved. Our position is that it is rather unreasonable
to expect investigations that are motivated by school students themselves
(an established element of good practice in independent research projects,
\citealp[e.g.][]{dunlop19}) to be able to make meaningful contributions
to the physics research as a matter of course. We note that in some
exceptional cases PRiSE students' work has arrived at promising preliminary
results, though these have required significant follow-up work by
professional researchers to transform the results into publishable
research \citep[e.g.][]{archer18} and thus should not be considered
the archetype. Instead, researchers are enticed by the possibility
of societal impact underpinned by their research. This is something
which is increasingly called upon from funders \citep[e.g.][]{nccperef}
and is notoriously difficult for areas of `blue skies' research such
as physics. Furthermore, significant contribution to a coordinated
departmental outreach programme can be used as criteria for academic
promotions \citep[cf.][]{hillier19}. Physics researchers though are
largely unmotivated in delivering curriculum content as part of their
engagement work, valuing instead aspects relating to their research
and role as a researcher \citep{thorley16}. PRiSE thus also aligns
with this direction. Ultimately, researcher buy-in is vital to the
delivery of protracted research-based engagement programmes such as
PRiSE.

It is clear from Table~\ref{tab:projects} that the topics and activities
of current PRiSE projects vary considerably. This suggests that a
wide range of fields and project ideas might be able to adopt the
PRiSE framework. How projects have been developed has also varied
(further explored in the supplementary material) though we have adopted
a pragmatic approach in taking advantage of opportunities (grant funding,
internships etc.) and adapting existing materials where possible,
since creating a project from scratch is a significant undertaking
far beyond what most academics (unfortunately) have capacity to do
\citep[cf.][]{thorley16}.

\subsection{Structure\label{subsec:Structure}}

\begin{sidewaysfigure*}
\includegraphics[width=1\columnwidth]{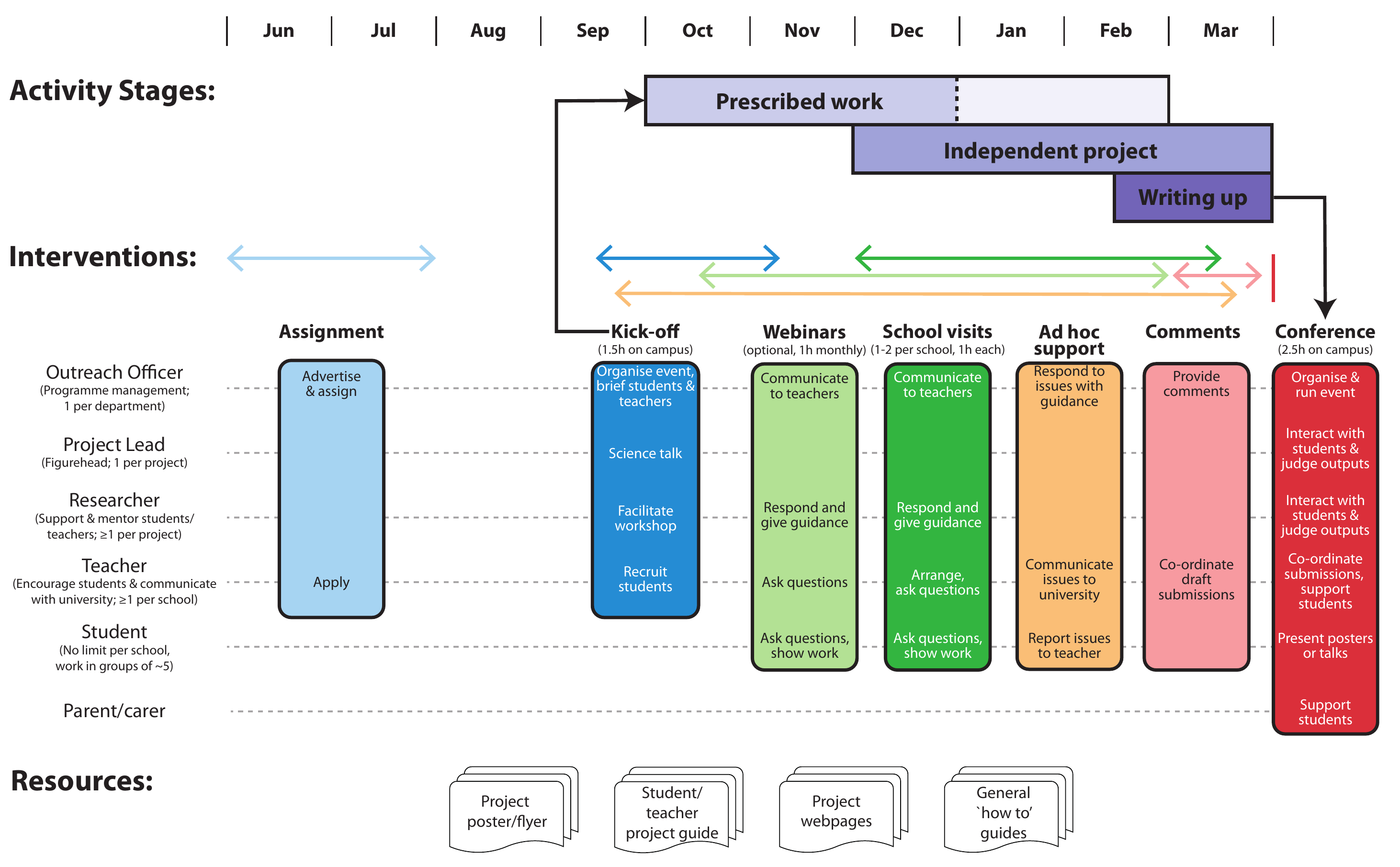}

\caption{Graphic summarising the PRiSE framework including a timeline of the
different project activity stages (rectangles), interventions and
stakeholders' roles within them (rounded rectangles), and resources
provided (document shapes). Arrows indicate over what dates interventions
(identified by colour) typically occur.\label{fig:framework}}
\end{sidewaysfigure*}

PRiSE runs from the start of the UK academic year to just before the
spring/Easter break, which teachers had informed us during the pilot
stage is manageable and largely fits around exams / other activities
for most (but not necessarily all) schools \citep{archer_report17}.
The structure has evolved naturally from the pilot to that shown in
Figure~\ref{fig:framework}.

\subsubsection{Activity stages}

Students work in research groups of typically five people and they
are advised to try and work on the project on average for 1--2 hours
a week. The bulk of this is done outside of regular physics lessons,
though some schools integrate the projects within their timetabled
`science clubs' or required extra-curricular blocks, whereas other
teachers arrange a regular slot for students to work on the projects
or leave it up to the students to arrange (though this latter approach
often proves unsuccessful). PRiSE projects are split up into three
activity stages (see Figure~\ref{fig:framework}).
\begin{itemize}
\item \textbf{Prescribed work:} Given that independent research in STEM
is probably unfamiliar to the students, rather than expecting them
to be able to come up with their own avenues of investigation in an
unfamiliar research topic straight away, we instead give them an initial
prescribed stage of research.
\item \textbf{Independent project: }Groups are encouraged to set their own
research questions and undertake different projects in the topic area
when ready. This enables every group to explore something different
so that students gain a sense of independence and ownership of their
work.
\item \textbf{Writing up: }Near the end of the project students produce
either a scientific poster or talk to be presented at an annual conference.
\end{itemize}

\subsubsection{Interventions}

\begin{figure*}
\includegraphics[width=1\columnwidth]{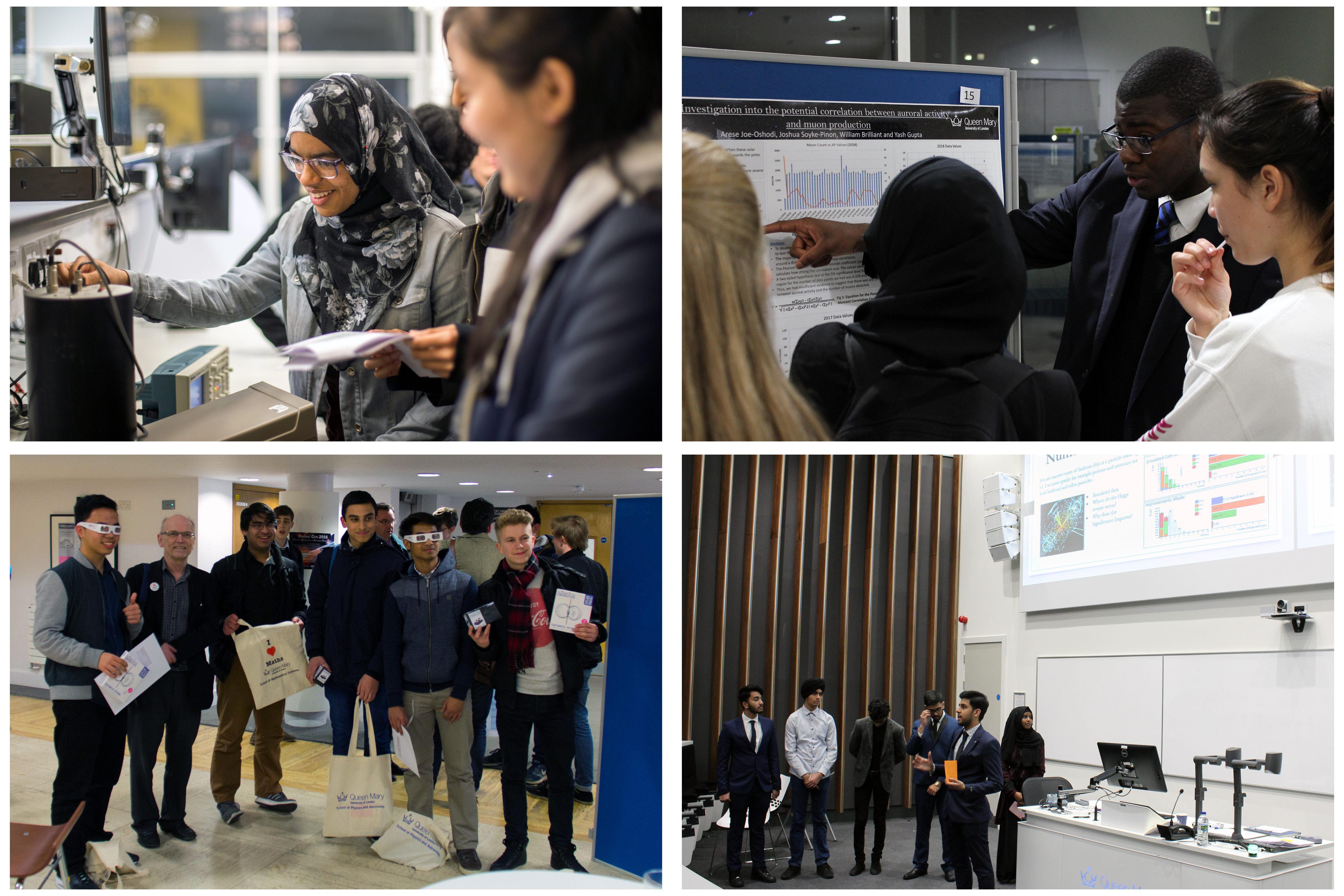}\caption{Photos of various stages of the PRiSE framework: students participating
in an on campus kick-off workshop (top left), students interacting
during the poster session at a conference (top right), a group of
students display their prizes won at a conference along with their
teacher (bottom left), a group present a talk at a conference (bottom
right).\label{fig:photos}}
\end{figure*}

Several different interventions form the structure and support behind
the activity stages:
\begin{itemize}
\item \textbf{Assignment:} The opportunity is advertised to schools via
existing teacher networks and teachers apply to participate in the
following academic year. Schools are assigned a project before the
summer break.
\item \textbf{Kick-off: }Typically on-campus event featuring an introductory
science talk, outline of how the project will work, and a hands-on
workshop.
\item \textbf{Visit: }Researchers visit schools to mentor students (and
their teachers) on their project work in a student-driven intervention.
\item \textbf{Webinars:} Drop-in online sessions similar to school visits
but allowing students and teachers to gain further support.
\item \textbf{Ad hoc: }Further asynchronous support via email as required.
\item \textbf{Comments:} Students are offered the opportunity to receive
comments on their draft slides or posters..
\item \textbf{Conference:} Students present the results of their projects
as oral or poster presentations at a special conference at the university,
attended by researchers as well as the students' teachers, peers,
and family.
\end{itemize}
Stakeholders' roles within these interventions are given in Figure~\ref{fig:framework},
with photos depicting some of them displayed in Figure~\ref{fig:photos}.
All stages of the programme and the processes involved are communicated
to teachers via email to pass on to their students. 

\subsubsection{Resources}

To enable the students to take part in PRiSE, the students and teachers
are also provided with numerous resources. While some projects require
specific equipment, data and software, here we discuss more common
types of resources across the different projects as shown in Figure~\ref{fig:framework}.
\begin{itemize}
\item \textbf{Project poster/flyer:} Given to teachers to help advertise
projects in their school.
\item \textbf{Project guide:} Each project has a student guide covering
an introduction to the research field, background physics/theory,
an explanation of the equipment/data, discussion of analysis techniques,
details of the initial prescribed activity, suggested research questions
/ methods for independent research, and links to other sources of
information. Teachers are provided with the same guide, but with extra
guidance.
\item \textbf{Project webpage:} These showcase anonymised examples of good
quality talks/posters that previous students have produced as well
as providing any links or videos relevant to the project.
\item \textbf{`How to' guide:} General articles applicable to most research
projects, such as producing scientific talks and posters.
\end{itemize}

\subsection{Scalability}

PRiSE's framework attempts to find a balance between the (necessarily
competing) reach and significance of the interactions. For example,
an academic staff member acting as both project lead and researcher
within PRiSE can support 4~schools' participation ($\sim$50~students),
taking around 8~hours over the course of 6~months (cf. Figure~\ref{fig:framework}).
Using PhD students or post-docs in the researcher role(s) makes even
more efficient use of time. In contrast, under the ORBYTS model each
early career researcher can support only one school (4--6~students)
with 10~hours of their time. As noted in the introduction, mentorship
from active researchers throughout does not appear prevalent in other
`research in schools' programmes at present.

Programmes of repeat-interventions with schools will necessarily have
a smaller reach than various one-off events. However, PRiSE's efficiency
in researcher time is reflected in its reach as shown in Table~\ref{tab:growth},
demonstrating the model now serves around 30 schools per year having
reached a total of 67~schools and over 1,300~students with the direct
involvement of 88~teachers as of 2020. We note that a minority of
schools do not complete the full programme and others do not return
for subsequent years, however, we do not provide a full analysis of
the types of schools involved and their retention within the programme
in this paper. Comparing the number of schools to other physical science
`research in schools' projects/programmes: University of Oxford researchers
have interacted directly with only 14~students from 5~schools through
their Higgs Hunters IRIS project which commenced in 2016 (O. Moore,
personal communication, 2020; though we note information regarding
other IRIS projects is not available); \citet{orbyts} reports collaborating
with 17~schools since 2016; and \citet{hisparc} discloses 140 schools
since 2004 across their network of 13 Dutch, Danish, and UK universities
\citep{vandam20}, i.e. around 11 per institution. Therefore the reach
through the PRiSE framework by a single university department is considerable
compared to the rest of the sector.

\begin{table*}
\begin{tabular}{cr|cccccc}
\multicolumn{2}{r|}{Academic Year} & 2014/2015 & 2015/2016 & 2016/2017 & 2017/2018 & 2018/2019 & 2019/2020\tabularnewline
\hline 
\multirow{3}{*}{Number per year} & Schools & 1 & 6 & 18 & 29 & 27 & 33\tabularnewline
 & Students & 20 & 115 & 163 & 310 & 311 & 407\tabularnewline
 & Teachers & 1 & 7 & 25 & 29 & 31 & 38\tabularnewline
\hline 
\multirow{3}{*}{Unique cumulative total} & Schools & 1 & 6 & 20 & 39 & 50 & 67\tabularnewline
 & Students & 20 & 135 & 298 & 608 & 919 & 1326\tabularnewline
 & Teachers & 1 & 7 & 28 & 44 & 63 & 88\tabularnewline
\end{tabular}

\caption{Number of schools, students and teachers involved in PRiSE by academic
year as the programme has grown.\label{tab:growth}}
\end{table*}

\section{Methods}

To determine the perceived value of PRiSE's approach with its key
stakeholders, namely participating students and teachers as well as
those across the wider university sector, we have maintained regular
collection of evaluative data \citep[cf.][and references therein]{rogers14}
via various surveys which we detail here. This data underpins our
understanding of PRiSE and has been collected securely to protect
all participants, in compliance with GDPR and in line with the \citet{bera18}
guidelines for educational research.

\subsection{Instruments and participants}

We gathered feedback from participating students and teachers via
paper questionnaires handed out at our student conferences each year.
The only exception to this was in 2020, where online forms were used
due to the COVID-19 pandemic causing that year's conference to be
postponed. The questionnaire method was chosen so as to gather data
from as wide a range of students and teachers as possible, respecting
the limited time/resources of all involved (both on the school and
university sides). For ethics considerations all feedback was anonymous,
with students and teachers only indicating their school (pseudonyms
are used here to protect anonymity) and which project they were involved
with. Students were not asked to provide details of any protected
characteristics (such as gender or race) or sensitive information
(such as socio-economic background). Both students and teachers were
informed via an ethics statement on the form that the information
was being collected for the purpose of evaluating and improving the
programme and that they could leave any question they felt uncomfortable
answering blank (this functionality was also implemented on the online
form for consistency).

The open and closed questions concerning participants' experience
of the programme, which varied slightly year-to-year, are given in
Appendix~\ref{sec:questions}. While we attempted to collect responses
from all participants in attendance, invariably only a fraction did
so yielding results from 153~students and 45~teachers across 37~schools.
A breakdown of the number of respondents and their schools per year
is given in Table~\ref{tab:response-rates}, where the number of
participants and schools in attendance at our conferences are also
indicated. We do not have reliable information on how many students,
teachers, and schools would have successfully completed the programme
in 2020 due to the COVID-19 disruption. Students and teachers did
not always answer all of the questions asked, hence we indicate the
number of responses for each question considered throughout. There
is no indication that the respondents differed in any substantive
way from the wider cohorts participating in the programme. While ideally
one would also gather feedback from schools which dropped out during
the year, a similar formal feedback process has not been viable bar
in a few cases where only the teachers responded.

\begin{table}
\begin{centering}
\begin{tabular}{c|cccccc}
 & \textbf{2015} & \textbf{2016} & \textbf{2017} & \textbf{2018} & \textbf{2019} & \textbf{2020}\tabularnewline
\hline 
\textbf{Students} &  & 13/26 (50\%) & 21/70 (30\%) & 46/92 (50\%) & 38/97 (39\%) & 35/?\tabularnewline
\textbf{Teachers} & 1/1 (100\%) & 6/6 (100\%) & 6/11 (55\%) & 9/16 (56\%) & 6/16 (38\%) & 17/?\tabularnewline
\textbf{Schools} & 1/1 (100\%) & 6/6 (100\%) & 11/ 11 (100\%) & 13/15 (87\%) & 11/15 (73\%) & 19/?\tabularnewline
\end{tabular}
\par\end{centering}
\caption{Response rates to questionnaires at PRiSE student conferences.\label{tab:response-rates}}
\end{table}

Feedback from the university sector came from a session at the 2019
Interact symposium that presented the challenges to STEM outreach
practice highlighted by recent educational research, the need for
deeper programmes of engagement with young people, and then summarised
the PRiSE approach as one possible example \citep{Archer2019interact}.
Throughout this workshop an anonymous interactive online survey was
used for interactivity and to collect data presented in this paper.
The survey included both closed and open questions as listed in Appendix~\ref{sec:University-sector-questionnaire}.
Attendees were fairly evenly split between UK university researchers
and engagement professionals (gauged in-person by attendees raising
their hands when asked), with 19 people participating in the survey
and only 7 not doing so. Participants were allocated a unique number
by the online survey itself, which did not distinguish between researchers
and engagement professionals.

\subsection{Analysis}

Both qualitative and quantitative approaches were utilised in data
analysis, as the open and closed ended questions present in the questionnaires
produced different types of data.

For all quantitative data, standard (i.e. 68\%) confidence intervals
are presented throughout. For proportions/probabilities these are
determined through the \citet{clopper34} method, a conservative estimate
based on the exact expression for the binomial distribution, and therefore
represent the expected variance due to counting statistics only. Several
statistical hypothesis tests are used with effect sizes and two-tailed
$p$-values being quoted, with the required significance level being
$\alpha=0.05$. In general we opt to use nonparametric tests as these
are more conservative and suffer from fewer assumptions (e.g. normality,
interval-scaling) than their parametric equivalents such as t-tests
\citep{hollander99,gibbons11}. The Wilcoxon signed-rank test is used
to compare single samples to a hypothetical value, testing whether
differences in the data are symmetric about zero in rank. When comparing
unpaired samples a Wilcoxon rank-sum test is used, which tests whether
one sample is stochastically greater than the other (often interpreted
as a difference in medians). Finally, for proportions we use a binomial
test, an exact test based on the binomial distribution of whether
a sample proportion is different from a hypothesized value \citep{howell07}.
For ease of reference, further details about the quantitative analyses
are incorporated into the relevant sections of the findings.

Qualitative data were analysed using thematic analysis \citep{Braun2006}.
Instead of using a priori codes, the themes were allowed to emerge
naturally from the data using a grounded theory approach \citep{Robson2011,Silverman2010}
as follows:
\begin{enumerate}
\item Familiarisation: Responses are read and initial thoughts noted.
\item Induction: Initial codes are generated based on review of the data.
\item Thematic Review: Codes are used to generate themes and identify associated
data.
\item Application: Codes are reviewed through application to the full data
set.
\item Analysis: Thematic overview of the data is confirmed, with examples
chosen from the data to highlight the themes.
\end{enumerate}

\section{Feedback from participants\label{sec:Feedback}}

In this section we use the feedback from participating students and
teachers to evaluate the provision offered within the PRiSE framework,
specifically assessing their experience and the level of support offered.

\subsection{Experience}

\begin{figure}
\begin{centering}
\includegraphics{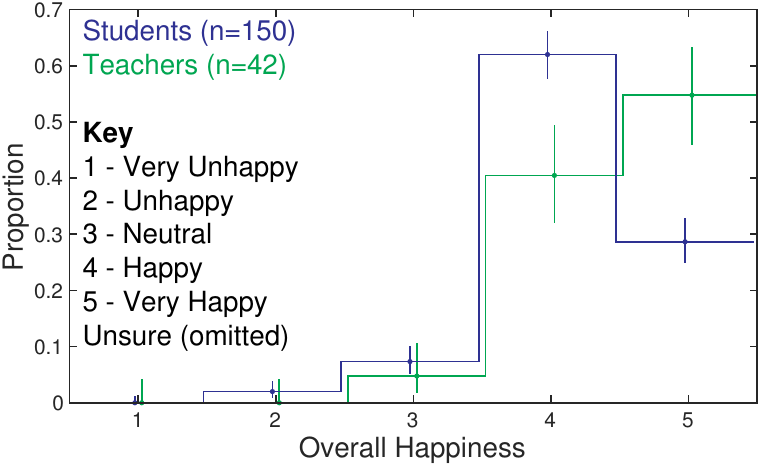}
\par\end{centering}
\caption{Distribution of students' (blue) and teachers' (green) overall happiness
with their PRiSE projects. Error bars denote standard ($1\sigma$)
\citet{clopper34} intervals.\label{fig:happiness}}
\end{figure}

Firstly from 2016 onwards we asked both students ($n=150$) and teachers
($n=42$) ``Have you been happy with the research project overall?''
giving options on a 5-point Likert scale, which we coded to the values
1--5. This scale and the results are displayed in Figure~\ref{fig:happiness},
revealing that $91\pm3\%$ of students and $95\pm5\%$ of teachers
rated their experience as positive (scores of 4--5) with only three
students giving a negative reaction (scores of 2). Teachers tended
to rank this question somewhat higher (their mean score was $4.50\pm0.09$,
where uncertainties refer to the standard error in the mean) than
students (mean of $4.17\pm0.05$), with $p=0.002$ in a Wilcoxon rank-sum
test. The PHwP project scored slightly higher (average of $4.59\pm0.11$,
$p=8\times10^{-4}$) than the overall results with students, whereas
ATLAS scored slightly lower with both students ($3.92\pm0.09$, $p=0.012$)
and teachers ($3.80\pm0.20$, $p=0.017$) than their respective means.
No obvious trends were present by school.

While suggestive of extremely positive experiences with PRiSE, one
also needs to compare these distributions against the typical responses
of students and teachers for schools STEM engagement programmes. We
use the results of \citet{vennis17} as such a benchmark, which surveyed
729 high-school students and 35~teachers about 12 different STEM
outreach activities in the USA and Netherlands. This comparison reveals
that PRiSE seems to be perceived considerably more positively than
usual by both students (benchmark average $3.66\pm0.01$, $p=1\times10^{-15}$
in a one-sample Wilcoxon signed-rank test) and teachers (benchmark
average $3.84\pm0.08$, $p=1\times10^{-7}$).

\begin{figure}
\begin{centering}
\includegraphics{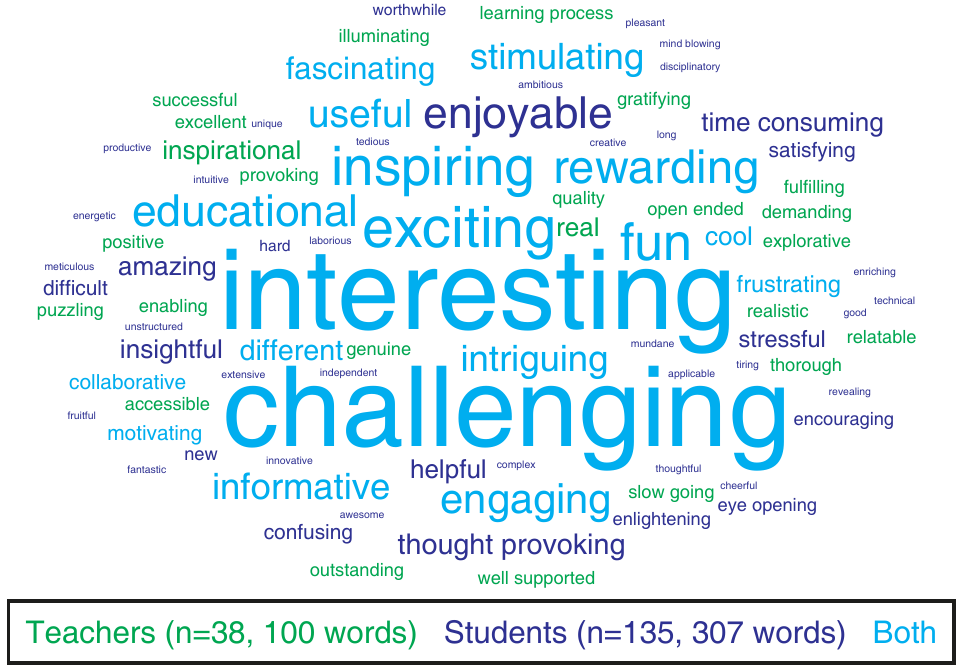}
\par\end{centering}
\caption{Word cloud of students' experiences. Colours indicate words identified
by students (blue), teachers (green), or both (cyan). Students and
teachers have been given equal total weight.\label{fig:keywords}}
\end{figure}

Secondly, students ($n=135$) were asked for adjectives describing
their experience of the projects overall. They were free to use any
words they wanted and were not given a pre-selected list. Teachers
($n=38$) were similarly asked to indicate observations of their students'
experience also. Since 2016 this has resulted in 88 unique adjectives,
with both students and teachers typically writing 2--3 words each.
We present the results as the word cloud in Figure~\ref{fig:keywords},
where students and teachers have been given equal prominence by normalising
their counts by their respective totals. We have indicated by colour
from which group(s) the words originated, generally showing a lot
of agreement between students' thoughts and teachers' observations.
The most cited adjectives were (in descending order) interesting,
challenging, exciting, inspiring and fun, similar to those from the
pilot \citep{archer_report17}, with the top two adjectives being
significantly greater than the subsequent ones. While in the pilot
stage only positive adjectives were expressed, since then a few negative
experiences have been conveyed such as time consuming, frustrating
and stressful. These constitute a small minority of experiences though
($6\pm1\%$) and in most cases the same students also listed positive
adjectives apart from only four individuals.

Following on from these quantitative analyses, we qualitatively explore
the potential reasons behind the results. The most common themes that
emerged from students' ($n=110$) responses to open questions about
their experience were that they feel they learnt a lot (62~responses)\begin{quote}\textit{
``We have learnt so many new things relating to the magnetosphere
and waves and we have developed new skills.}'' (Student~3, Xavier's
Institute for Higher Learning, MUSICS 2016)\\
``\textit{I learned so much! I would recommend it to all the younger
kids at my school!}'' (Student 111, St Trinians, SCREAM 2019)\\
``\textit{I definitely learnt many new and interesting things and
it helped me to develop my understanding of particle physics while
aiding my A-Level knowledge.}'' (Student~140, Jedi Academy, ATLAS
2020);\end{quote}they found the projects' content and methods interesting
(47~responses)\begin{quote}``\textit{It brought my interests in
programming, physics, maths and space together}.'' (Student~100,
Octavian Country Day School, PHwP 2019)\\
``\textit{It has been very interesting to work with actual data and
plan our own research project}.'' (Student~116, St Trinians, SCREAM
2019)\\
``\textit{It was very interesting to learn more in regards to astrophysics
and the MUSICS project was a very safe space to do so. We got lots
of support and it was fun.}'' (Student~119, Pok\'{e}mon Technical
Institute, MUSICS 2020)\\
``\textit{Coming into the Planet Hunting With Python project, my
interests were mainly focused on the physics side of understanding
brightness-curves and finding equations to solve for planetary parameters.
However, in this project, my eyes were opened to the many uses of
coding to analyse data, and it was a wonderful experience to learn
about such an interesting area through a combination of theory and
practical coding.}'' (Student~120, Octavian Country Day School,
PHwP 2020);\end{quote}and they enjoyed the style of working in and
with research which differs from their regular school experience (33~responses)\begin{quote}``\textit{It
was quite interesting to broaden our views and experience high level
education.}'' (Student~90, Hill Valley High School, ATLAS 2019)\\
``\textit{I enjoyed the opportunity to do science instead of just
learning it.}'' (Student~101, Octavian Country Day School, PHwP
2019)\\
``\textit{It was very nice to work with friends and work together
to produce something.}'' (Student~106, Boston Bay College, PHwP
2019)\\
``\textit{It was very fun to do our own research and I appreciated
that help was always available even thought it is very independent.
It also shows how challenging research can actually be but also how
rewarding it is once you start making progress.}'' (Student~143,
Sunnydale High School, MUSICS 2020)\end{quote}Neutral or negative
experiences tended to be due to students finding the projects' content
or open-ended way of working difficult or confusing (7~responses)\begin{quote}``\textit{I
did not understand most of the project or what I was supposed to do.}''
(Student~147, Jedi Academy, ATLAS 2020)\end{quote}The vast majority
of students seemed to ultimately enjoy this challenge though\begin{quote}``\textit{The
project gave us a lot of freedom and challenged us to think in different
ways.}'' (Student~122, Jedi Academy, ATLAS 2020)\\
``{[}It{]} \textit{made me more willing to take a go at challenges
and what I deem hard.}'' (Student~130, Bending State College, PHwP
2020)\end{quote}and teachers agreed that the learning curve involved
with the projects was advantageous for students\begin{quote}``\textit{When
the students got the hang of it they really enjoyed it.}'' (Teacher~16,
Tree Hill High School, MUSICS 2018)\\
``\textit{The students found it hard to identify what to do a project
on and would have liked guidance on that, but I felt this was a good
experience. They have developed grappling with open ended and difficult
material.}'' (Teacher~23, Smeltings, ATLAS 2019)\end{quote}Teachers'
feedback on their experience ($n=34$) tended to praise how the projects
allow their students to access and explore beyond the curriculum (29~responses)\begin{quote}``\textit{A
great framework to explore physics beyond the syllabus but still accessible.}''
(Teacher~26, Octavian Country Day School, PHwP 2019)\\
``\textit{Excellent project that is open-ended allowing students
to take it where they want.}'' (Teacher~27, Hogwarts, SCREAM 2019)\\
``\textit{This year I had an extremely motivated, enthusiastic and
well-organised group of 7 students who fully immersed themselves into
the project and quickly took it in a direction outside my own understanding
of this area of science. This is exactly the experience I wanted them
to have, and they were able to discover some genuinely novel processes
that had not been observed before - the hallmark of great scientific
research!}'' (Teacher~44, Sunnydale High School, MUSICS 2020)\end{quote}Therefore,
both quantitative and qualitative data suggest students and teachers
had much more positive and rewarding experiences participating in
PRiSE projects than is typical for schools engagement programmes from
universities in general.

\subsection{Support and resources}

We originally asked students whether they felt they had received adequate
support, finding overall positive results on a 5-point Likert scale
\citep{archer_report17}. However, students' qualitative responses
explaining their answers often revealed a conflation of the support
provided by Queen Mary with that offered by their teacher. Therefore,
from 2019 onwards we explicitly separated these two aspects. Students
($n=68$) were asked ``Do you feel that support from your teacher
was provided/available during the project?'' which yielded the following
results: Strongly Agree (30), Agree (34), Neither Agree or Disagree
(3), and Disagree (1). The average response is $4.37\pm0.08$, which
is considerably greater than the benchmark on teacher support reported
by \citet{vennis17} of $3.60\pm0.03$ ($p=4\times10^{-10}$). Students'
comments explaining their ratings ($n=56$) revealed that teachers
provided them with advice, encouragement, and enthusiam (49~responses)\begin{quote}
``\textit{My teacher has been very supportive and has helped us when
we didn't understand something as well as encouraging us to taking
a more innovative approach.}'' (Student~124, Quirm College for Young
Ladies, MUSICS 2020)\\
``\textit{If we had a question, teachers were probably not useful.
But if we did not know what to do or we were stuck, here teachers
were really useful and that was what we needed.}'' (Student~145,
Sunnydale High School, MUSICS 2020)\end{quote}as well as arranging
regular sessions for students to meet and visits or calls from the
university when required (7~responses)\begin{quote}``\textit{Our
teacher arranged a Skype call with a professor from QMUL when we needed
to ask questions about how certain parts of the data were calculated.}''
(Student~126, Harbor School, ATLAS 2020)\\
``{[}Our{]} \textit{teacher would often ask us about it and hold
meetings to catch up with us on our progress.}'' (Student~143, Sunnydale
High School, MUSICS 2020)\\
``{[}Our{]} \textit{teacher answered some of our questions and organised
a weekly meet-up where students could ask each other questions and
work together.}'' (Student~148, Jedi Academy, ATLAS 2020)\end{quote}Neutral
or negative responses tended to be explained by their teacher lacking
specific knowledge about the research in response to students' queries
(2~responses)\begin{quote} ``\textit{Although they were always
ready to help, sometimes they didn't know the answers to our questions.}''
(Student~122, Jedi Academy, ATLAS 2020)\\
``{[}They{]} \textit{didn't understand the content of the project.}''
(Student~139, Jedi Academy, ATLAS 2020)\end{quote}which is something
we don't expect of teachers \citep[cf.][]{shah16,bennett16,bennett18},
hence why support from the university is also offered.

Teachers' ($n=18$) responses on a yes/no scale (chosen due to expected
small number statistics) of whether they felt able to support their
students were also highly positive with only 2 negative responses,
a significant majority ($p=0.001$ in a two-tailed binomial test).
Bear in mind, however, that these responses were in light of the support
provided from the university, something which a few teachers referenced
in explaining their answers\begin{quote} ``\textit{My own experience
with research was handy but I felt that without this the students
would still have been supported}'' (Teacher~2, Hogwarts, SCREAM
2016)\\
``\textit{only through Martin's support}'' (Teacher~4, Sweet Valley
High School, MUSICS 2016)\\
``\textit{the teacher version of the handout was useful, but otherwise
I could only give generic advice}'' (Teacher~16, Tree Hill High
School, MUSICS 2018)\end{quote}Several teachers raised pressure on
their time, with one teacher using this to justify their negative
response\begin{quote}``\textit{sheer time pressure - big limiting
factor}'' (Teacher~13, St Trinians, SCREAM 2017)\end{quote}another
quoting this somewhat limited their ability to support students\begin{quote}``\textit{not
as much as I would have liked (lack of available time)}'' (Teacher~3,
Xavier's Institute for Higher Learning , MUSICS 2016)\end{quote}but
most saying it was manageable\begin{quote}``\textit{The autonomous
group work, with very little input from me, was great to see}'' (Teacher~1,
Hogwarts, SCREAM 2015)\\
``\textit{Students were quite self-sufficient so if I made suggestions
they were able to do the leg work}'' (Teacher~15, Spence Academy
for Young Ladies, MUSICS 2018)\end{quote}Teachers' ability and confidence
in supporting the projects was another theme that emerged. Even with
the teacher-specific resources provided, some felt they did not have
the specific knowledge or skills to support the projects\begin{quote}``{[}Unable
to support due to a{]}\textit{ lack of knowledge of Python}'' (Teacher~19,
Boston Bay College, PHwP 2018)\end{quote}Other teachers reflected
that, similarly to their students, they too had experienced a learning
curve through their involvement \begin{quote}``{[}I found it{]}\textit{
difficult at first}'' (Teacher~8, Coal Hill School, MUSICS 2017)\end{quote}ultimately
becoming more determined and confident with time and in subsequent
years \begin{quote}``\textit{First time we've done this --- I will
do better next time}'' (Teacher~17, Sunnydale High School, MUSICS
2018)\\
``\textit{Second year that I ran it I feel more confident}'' (Teacher~21,
Hogwarts, SCREAM 2018)\end{quote}The final theme raised was that
for successful participation teachers believed the students needed
the external motivation coming from the university rather than having
project delivery being solely teacher-driven\begin{quote}``\textit{Dr
Archer was a great external lead to have. If I had been pushing them
myself they would have taken it less seriously}'' (Teacher~17, Sunnydale
High School, MUSICS 2018)\end{quote}Therefore, the comments from
both students and teachers indicate that teachers alone would likely
not have been able to successfully support these research projects
in their schools without both the resources and external motivation/mentoring
provided as part of the PRiSE framework.

\begin{figure}
\begin{centering}
\includegraphics{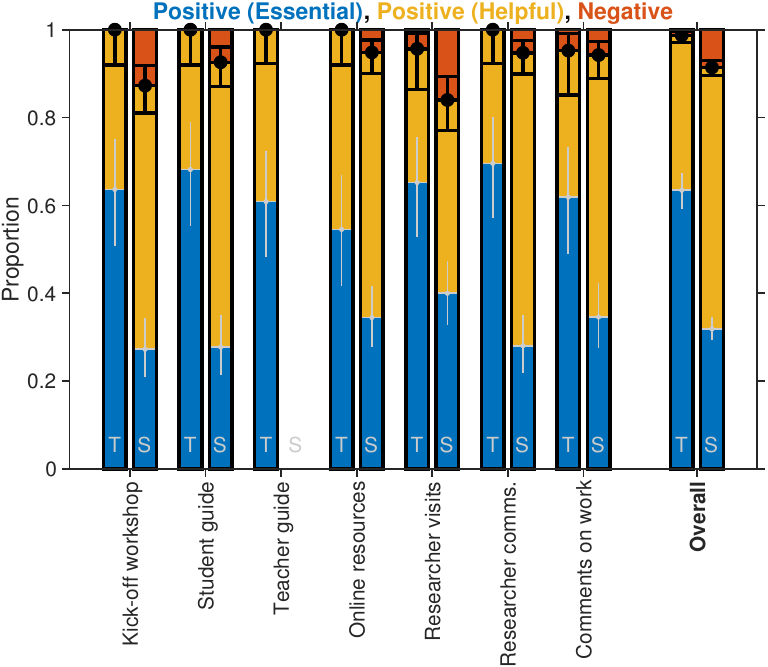}
\par\end{centering}
\caption{Usefulness of support provided to teachers (T, $n=23$) and students
(S, $n=68$). Results are divided (black lines and associated error
bars) into negative (red) and positive responses, with the latter
subdivided (grey lines and error bars) into `essential' (blue) and
`helpful' (yellow) elements. Error bars denote standard ($1\sigma$)
\citet{clopper34} intervals. \label{fig:support}}
\end{figure}

We now consider the specific elements of support shown in Figure~\ref{fig:framework}.
From 2019 onwards we investigated participants' thoughts on each of
these various aspects offered. Students ($n=68$) and teachers ($n=23$)
were asked to rate the usefulness of these as either `unimportant',
`helpful', `essential', or `unsure'. This was chosen over a 5-point
Likert scale due to an expected low number of responses, particularly
from teachers. Any unsure or blank responses are neglected yielding
326 (out of a potential 408) student and 156 (out of 161) teacher
responses. We divide these responses into negatives (`unimportant')
and positives (`helpful' or `essential'), though we acknowledge some
may consider the `helpful' response as neutral and thus our analysis
takes both interpretations into account. The results are displayed
in Figure~\ref{fig:support} for the individual elements as well
as overall results obtained from totalling all responses. Both students
and teachers overall rated the elements positively --- coding the
responses to values of 1 (negative) to 3 (essential) the overall means
were $2.62\pm0.04$ for teachers and $2.23\pm0.03$ for students.
The majority of teachers tended to give `essential' ratings to most
aspects and while these majorities are not statistically significant
in a two-tailed binomial test, the average value for each element
was greater than 2 to high confidence ($p<0.002$ in one-sample Wilcoxon
signed-rank tests). Students, on the other hand, mostly rated each
element as `helpful' as well as stating slightly more negative responses
than teachers, though again all elements' mean scores (apart from
the kick-off workshop at $2.15\pm0.08$) were significantly greater
than 2 ($p<0.023$). While there are some variations in scoring amongst
the different support elements, such as students and teachers respectively
rating researcher visits and communications (the latter of which includes
webinars and ad hoc emails) as the most essential, these differences
to each group's overall results are slight and not statistically significant.
One interpretation of this might be that most respondents answered
unreflectively, ticking the same boxes for each item. However, no
students and only 3~teachers gave the same answer in every category.
This therefore suggests that all of the elements of support provided
as part of PRiSE are almost equally important and necessary. This
has been further elaborated on in teacher feedback:\begin{quote}``\textit{It
is very well set up and open ended and the support received is magnificent.}''
(Teacher~33, Boston Bay College, PHwP 2020)\\
``\textit{Martin's guide to help the students was a very good balance
of useful guidance and allowing them to find their own way through.}''
(Teacher~38, Royal Dominion College, MUSICS 2020)\\
``\textit{Truly excellent support from the Queen Mary team! They
have visited us multiple times and have been so generous with their
time. Students have learnt a great deal from them!}'' (Teacher~44,
Smallville High School, PHwP 2020)\end{quote}

\section{Feedback from the university sector\label{sec:Future}}

Based on the highly positive results from participants, we think there
is potential for the PRiSE framework to spread beyond QMUL and be
applied to other institutions' own areas of physics (and perhaps even
STEM more generally) research. We therefore wanted to assess how it
is perceived by those from the university sector with interests and/or
expertise in schools engagement. Feedback from our partner organisations
seemed promising, for example with the South East Physics Network
including PRiSE in their public engagement strategy \citep{sepnet},
several of their member institutions expressing interest in adopting
it, and the University of Surrey having already developed a pilot
project. However, for a slightly more balanced perspective we collected
evaluative data via an anonymous interactive online survey (see Appendix~\ref{sec:University-sector-questionnaire})
from independent researchers and public engagement professionals during
a session at Interact 2019 \citep{Archer2019interact}, a day-long
UK symposium concerning public engagement in the physical sciences.
This workshop contained an overview of recent educational research
and then introduced the PRiSE framework to attendees.

The respondents, while heavily bought into schools engagement, tended
to only undertake one-off activities (as detailed in Appendix~\ref{sec:University-sector-questionnaire}).
After presenting the PRiSE framework to them, when asked on a 5-point
Likert scale whether they ($n=19$) would now consider deeper approaches
to outreach / engagement with schools, the results were: Strongly
Agree (5), Agree (9), Neither Agree or Disagree (5), and no responses
in the two negative options. Coding these to a 1--5 scale yields
a mean of $4.0\pm0.2$, i.e. greater than neutral ($p=1\times10^{-4}$
in a one-sample Wilcoxon signed-rank test).

In an open-ended question, participants were also asked to identify
the main thing they had taken away from the session. This yielded
17~responses. Through coding these answers three main themes emerged,
with all responses covering only one theme each. The first theme is
about what types of schools engagement are possible (8~responses),
with almost all thinking that deeper programmes like PRiSE are achievable:\begin{quote}``\textit{Outreach
does not have to just be workshop/talk based. It can be an interactive
research based activity that supports research activities within HE}''
(Participant 24)\\
``\textit{students are probably far more capable than schools and
researchers might expect}'' (Participant~16)\end{quote}and only
one person claiming that such approaches are not practical \begin{quote}``\textit{Would
need a huge amount of time to set up something good - even with input
from other people!}'' (Participant~3)\end{quote}The second theme
(5~responses) concerns practical aspects towards delivering deeper
programmes based on PRiSE's approach:\begin{quote}``\textit{lots
of practical and multifaceted suggestions people in a variety of contexts
can take and adapt for themselves}'' (Participant~13)\\
``\textit{try and use existing resources available rather than reinventing
the wheel}'' (Participant~15)\end{quote}The final theme (5~responses)
concerned potential impacts from deeper programmes of schools engagement,
beyond the scope of this paper. These results suggest researchers
and engagement professionals also see the value in PRiSE's approach
and may be receptive to adopting the framework, though evidence of
action following these immediate attitudes is really needed. We also
acknowledge that this was a rather small survey from a group already
highly bought-in to schools engagement, thus results would likely
be less positive from a wider and more representative sample of all
researchers. These are avenues which could be explored further in
the future to gain a better perspective on whether the PRiSE framework
could realistically be rolled out further.

\conclusions{}

`Research in schools' programmes, where school students and their
teachers get to experience and interact with cutting-edge science
through independent projects over several months, may have some role
to play in addressing current issues around participation and equity
in science (physics in particular) at university. However, how best
to go about delivering such projects for schools remains unclear.
This paper evaluates the provision framework of the `Physics Research
in School Environments' (PRiSE) programme. These 6-month-long projects
mentored by expert researchers include a suite of activity stages,
interventions and resources to to enable a wide range of students,
teachers, and schools to be involved.

Feedback from participating students and teachers upon completion
has been significantly more positive than benchmark results on schools
engagement programmes with STEM in general. This is because students
and teachers have found the research projects of great interest and
have relished the challenge of working differently to their regular
school experience. This suggests that `research in schools' projects
are of greater value to schools than more common forms of outreach
activities. Since these experiences can uniquely be provided by active
researchers compared to other possible STEM engagement providers,
we strongly urge universities consider `research in schools' approaches
to engagement. Participants find the numerous elements of supporting
interventions and resources provided in PRiSE's framework, uncommon
in general with other schemes, as equally valued and necessary for
their participation. We therefore recommend that providers of independent
research projects offer schools detailed project guides, mentorship
from experts (in-person, live-online, and asynchronously), online
resources (such as `how to' articles, examples of previous students'
work, and videos), and finally the opportunity to present their work
to peers, teachers and family. We note that even with all the support
provided there is some attrition within the programme, which is to
be expected for any protracted engagement programme. While this has
not been explored here, it is assessed in a companion paper \citep{archer_prise_schools20}.
Currently though we have little data on the experience of students
and teachers that have dropped out of the programme, which is something
that could be explored in the future. We have also gathered data on
the impact of the PRiSE initiative, which is assessed elsewhere \citep{archer_prise_impact20}.

The PRiSE model attempts to make efficient use of researchers' time,
enabling more schools to be supported per institution than other current
formats. We have presented data from independent researchers and public
engagement professionals showing they seem receptive to the PRiSE
framework. Indeed, it is already slowly beginning to spread to other
institutions. This potential expansion might allow an assessment of
how generally applicable the framework is outside of its current London
location and what other affordances might be required in these contexts.
While PRiSE has so far only concerned areas of physics research, `research
in schools' in general already span all the sciences \citep[e.g.][]{bennett16,bennett18,iris20}.
We therefore see no reason why PRiSE's approach could not also be
broadened to other STEM areas, particularly areas of research based
in data and/or analysis. We therefore encourage researchers, and the
public engagement professionals who facilitate their activity, to
consider adopting this way of working and hope this paper can inform
this practice. In such cases, it is recommended that PRiSE projects
be embedded as core schools engagement activity within research groups.
We would be happy to support groups in developing, delivering, and
evaluating pilot PRiSE projects around their own research, thereby
making use of the learning that has arisen from the programe over
the last 6~years.

\appendix

\section{Information about UK/English schools\label{sec:UK-schools}}

To those unfamiliar with the UK/English education system system, we
provide some further notes here. The curriculum is broken down into
Key Stages of duration 2--4 years, with those for secondary schools
displayed in Table~\ref{tab:school-years}. The final two Key Stages
culminate in GCSE and A-Level examinations respectively, with the
latter being optional as education post-16 is not compulsory. Our
recommendations to teachers about which year groups should be involved
with PRiSE are also highlighted in the table.

\begin{table}
\begin{centering}
\begin{tabular}{|c|c|c|c|c|c|}
\hline 
\textbf{Key Stage} & \textbf{Year} & \textbf{Final Exam} & \textbf{Age} & \textbf{Policy} & \textbf{PRiSE involvement}\tabularnewline
\hline 
\multirow{3}{*}{KS3} & 7 & \multirow{4}{*}{None} & 11--12 & \multirow{5}{*}{Compulsory} & \multirow{3}{*}{Not recommended}\tabularnewline
\cline{2-2} \cline{4-4} 
 & 8 &  & 12--13 &  & \tabularnewline
\cline{2-2} \cline{4-4} 
 & 9 &  & 13--14 &  & \tabularnewline
\cline{1-2} \cline{2-2} \cline{4-4} \cline{6-6} 
\multirow{2}{*}{KS4} & 10 &  & 14--15 &  & \multirow{2}{*}{Select recommended}\tabularnewline
\cline{2-4} \cline{3-4} \cline{4-4} 
 & 11 & GCSE & 15--16 &  & \tabularnewline
\hline 
\multirow{2}{*}{KS5} & 12 & AS-Level (optional) & 16--17 & \multirow{2}{*}{Optional} & \multirow{2}{*}{All recommended}\tabularnewline
\cline{2-4} \cline{3-4} \cline{4-4} 
 & 13 & A-Level & 17--18 &  & \tabularnewline
\hline 
\end{tabular}
\par\end{centering}
\caption{Summary of the stages of secondary education in the English system.\label{tab:school-years}}
\end{table}

\section{Participant evaluation questions\label{sec:questions}}

Here we list the questions posed in questionnaires that are considered
within this paper, giving details of what phrasing was used, how participants
could respond, and which years the question was posed. Follow-on questions
are indicated by indentation and a down-right arrow (\rotatebox[origin=c]{180}{$\Lsh$}).
Students' responded to the following:
\begin{center}
\begin{footnotesize}%
\begin{tabular}{>{\raggedright}p{0.5\paperwidth}cc}
\textbf{Question} & \textbf{Response type} & \textbf{Year(s)}\tabularnewline
\hline 
\noalign{\vskip8pt}
Have you been happy with the research project overall? & 5-point Likert & 2016--2020\tabularnewline
\noalign{\vskip8pt}
\setlength{\leftskip}{1em}\rotatebox[origin=c]{180}{$\Lsh$}Please
tell us why / why not & Open Text & 2016--2020\tabularnewline
\noalign{\vskip8pt}
What adjectives would you use to describe your experience of the project
overall? & Open Keywords & 2016--2020\tabularnewline
\noalign{\vskip8pt}
Did you feel that support from your teacher was provided/available
during the project? & 5-point Likert & 2019--2020\tabularnewline
\noalign{\vskip8pt}
\setlength{\leftskip}{1em}\rotatebox[origin=c]{180}{$\Lsh$}Please
tell us why / why not & Open Text & 2019--2020\tabularnewline
\noalign{\vskip8pt}
Did you find the following elements of support from Queen Mary useful
in supporting you? & Closed Options & 2019--2020\tabularnewline
\noalign{\vskip8pt}
How could we improve future projects? & Open Text & 2016--2020\tabularnewline
\noalign{\vskip8pt}
Any other comments? & Open Text & 2016--2020\tabularnewline
\noalign{\vskip8pt}
\end{tabular}\end{footnotesize}
\par\end{center}

\noindent\begin{minipage}[t]{1\columnwidth}%
The questions asked of teachers were:
\begin{center}
\begin{footnotesize}%
\begin{tabular}{>{\raggedright}p{0.5\paperwidth}cc}
\centering{}\textbf{Question} & \textbf{Response type} & \textbf{Year(s)}\tabularnewline
\hline 
\noalign{\vskip8pt}
Have you been happy with the research project overall? & 5-point Likert & 2016--2020\tabularnewline
\noalign{\vskip8pt}
\setlength{\leftskip}{1em}\rotatebox[origin=c]{180}{$\Lsh$}Please
tell us why / why not & Open Text & 2016--2020\tabularnewline
\noalign{\vskip8pt}
What adjectives would you use to describe your students' experience
of the project overall? & Open Keywords & 2016--2020\tabularnewline
\noalign{\vskip8pt}
Did you feel able to provide support to your students during the project? & Yes/No & 2016--2018\tabularnewline
\noalign{\vskip8pt}
\setlength{\leftskip}{1em}\rotatebox[origin=c]{180}{$\Lsh$}Please
tell us why / why not & Open Text & 2016--2018\tabularnewline
\noalign{\vskip8pt}
Did you find the following elements of support from Queen Mary useful
in supporting you and your students? & Closed Options & 2019--2020\tabularnewline
\noalign{\vskip8pt}
\setlength{\leftskip}{1em}\rotatebox[origin=c]{180}{$\Lsh$}Are there
any ways we could further help you support your students with project
work? & Open Text & 2019--2020\tabularnewline
\noalign{\vskip8pt}
Would you be interested in running this project or a similar one with
us again? & Yes/No & 2016--2018\tabularnewline
\noalign{\vskip8pt}
\setlength{\leftskip}{1em}\rotatebox[origin=c]{180}{$\Lsh$}Please
tell us why / why not & Open Text & 2016--2018\tabularnewline
\noalign{\vskip8pt}
How could we improve future projects? & Open Text & 2016--2020\tabularnewline
\noalign{\vskip8pt}
Any other comments? & Open Text & 2016--2020\tabularnewline
\noalign{\vskip8pt}
\end{tabular}\end{footnotesize}
\par\end{center}%
\end{minipage}

\section{University sector questionnaire\label{sec:University-sector-questionnaire}}

The following questions were posed to university researchers and engagement
professionals via on online interactive survey during a session at
the 2019 Interact symposium \citep{Archer2019interact}.
\begin{center}
\begin{footnotesize}%
\begin{tabular}{lc}
\textbf{Question} & \textbf{Response type}\tabularnewline
\hline 
\noalign{\vskip8pt}
What do you want your outreach / engagement with schools to achieve
(i.e. what impact)? & Open Text\tabularnewline
\noalign{\vskip8pt}
What sorts of outreach activities with schools do you do? (Select
all that apply) & Closed Options\tabularnewline
\noalign{\vskip8pt}
I am now considering deeper approaches to outreach / engagement with
schools (Select only one) & 5-point Likert\tabularnewline
\noalign{\vskip8pt}
What is the main thing you have taken away from this session? & Open Text\tabularnewline
\noalign{\vskip8pt}
\end{tabular}\end{footnotesize}
\par\end{center}

\begin{figure}
\begin{centering}
\includegraphics{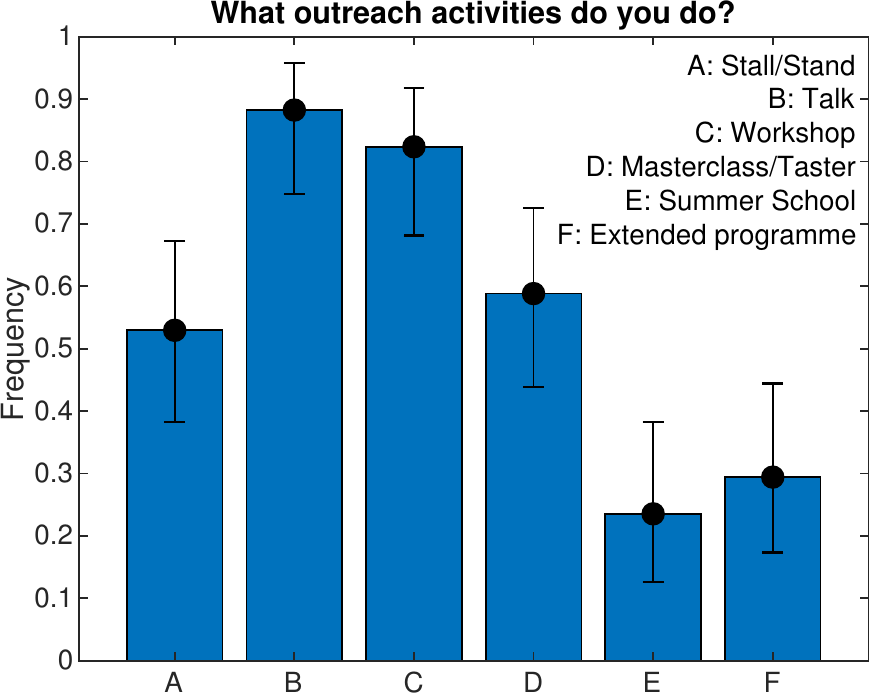}
\par\end{centering}
\caption{Bar chart of types of schools outreach activities undertaken by university
researchers and engagement professionals ($n=17$). Error bars denote
the standard ($1\sigma$) \citet{clopper34} interval.\label{fig:outreach-types}}
\end{figure}

For context on these participants, Figure~\ref{fig:outreach-types}
shows the distribution of the types of activity they undertake where
they could select from:\renewcommand{\labelenumi}{\Alph{enumi}.}
\begin{itemize}
\item Stall/stand: drop-in activities for schools at STEM or careers fairs
\item Talk: a typically one/two lesson slot featuring a predominantly one-way
interaction
\item Workshop: a typically one/two lesson slot with mostly two-way interaction
and often hands-on activities for students
\item Masterclass/taster: half-day or day-long activities which may be comprised
of talks and/or workshops
\item Summer school: several-day to week-long activities often involving
some project work as well as talks and/or workshops
\item Extended programme: multiple interventions with the same group of
students over a protracted period of time
\end{itemize}
\renewcommand{\labelenumi}{\arabic{enumi}.}Unsurprisingly, one-off
activities such as talks and workshops scored the highest whereas
more the protracted engagements, summer schools and extended programmes,
were significantly ($p<0.0019$) less common. The attendees were also
asked what they hoped to achieve (i.e. the aims or intended impacts)
through their school engagements via an open question. Performing
a thematic analysis of the qualitative results, it was possible to
categorise the majority of answers into the following:
\begin{itemize}
\item Change school students\textquoteright{} aspirations (9 people), with
the word \textquotedblleft inspire\textquotedblright{} often used
\item Enhance students\textquoteright{} awareness or understanding of STEM
(6 people), often in the context of primary research
\item Tackle societal biases in STEM (4 people), most often gender
\end{itemize}
Note that some responses covered more than one of these aims. Other
stated motivations outside of these themes included \textquotedblleft access
to a student population for {[}research{]} studies\textquotedblright ,
to \textquotedblleft build relationships\textquotedblright , and to
deliver \textquotedblleft meaningful content\textquotedblright . The
three themes are in general agreement with those determined by \citet{thorley16}
in a larger survey of UK physicists.

\dataavailability{Data supporting the findings of this study that
is not already contained within the article or derived from listed
public domain resources are available on request from the corresponding
author. This data is not publicly available due to ethical restrictions
based on the nature of this work.}

\authorcontribution{MOA conceived the programme and its evaluation,
performed the analysis, and wrote the paper. JDW and CT contributed
towards the analysis, validation, and writing. OK contributed to the
writing.}

\competinginterests{The authors declare that they have no conflict
of interest.}
\begin{acknowledgements}
We thank Dominic Galliano for helpful discussions. MOA and CT are
grateful for funding from the Ogden Trust. MOA holds a UKRI (STFC
/ EPSRC) Stephen Hawking Fellowship EP/T01735X/1. This programme has
been supported by a QMUL Centre for Public Engagement Large Award,
and STFC Public Engagement Small Award ST/N005457/1.
\end{acknowledgements}
\DeclareRobustCommand{\disambiguate}[3]{#1}\bibliographystyle{copernicus}
\bibliography{prise2020}

\end{document}


\renewcommand*\thesection{S\arabic{section}} 

\renewcommand*\thefigure{S\arabic{figure}} 

This supplementary material provides further details about the development
and implementation of the `Physics Research in School Environments'
(PRiSE) framework. It is aimed at practitioners interested in adopting
the practices at their own institutions.

\section{Aims}

There is no ``magic bullet'' to increasing physics (or more broadly
STEM) uptake and diversity at higher education --- multiple different
approaches are needed with each addressing different stages of young
people's educational journey as well as their key influencers and
wider learning ecology in relevant ways \citep[e.g.][]{davenport20}.
Furthermore, research has shown that young people's aspirations are
incredibly difficult to influence \citep{aspires13,archer14} with
standard one-off interventions, or even short-series, showing no real
changes, highlighting the need for more extended and in-depth programmes
for significant lasting impact \citep[see the review of][and references therein]{archer20r4a}. 

\begin{sidewaysfigure}
\centering{}\includegraphics[width=1\columnwidth]{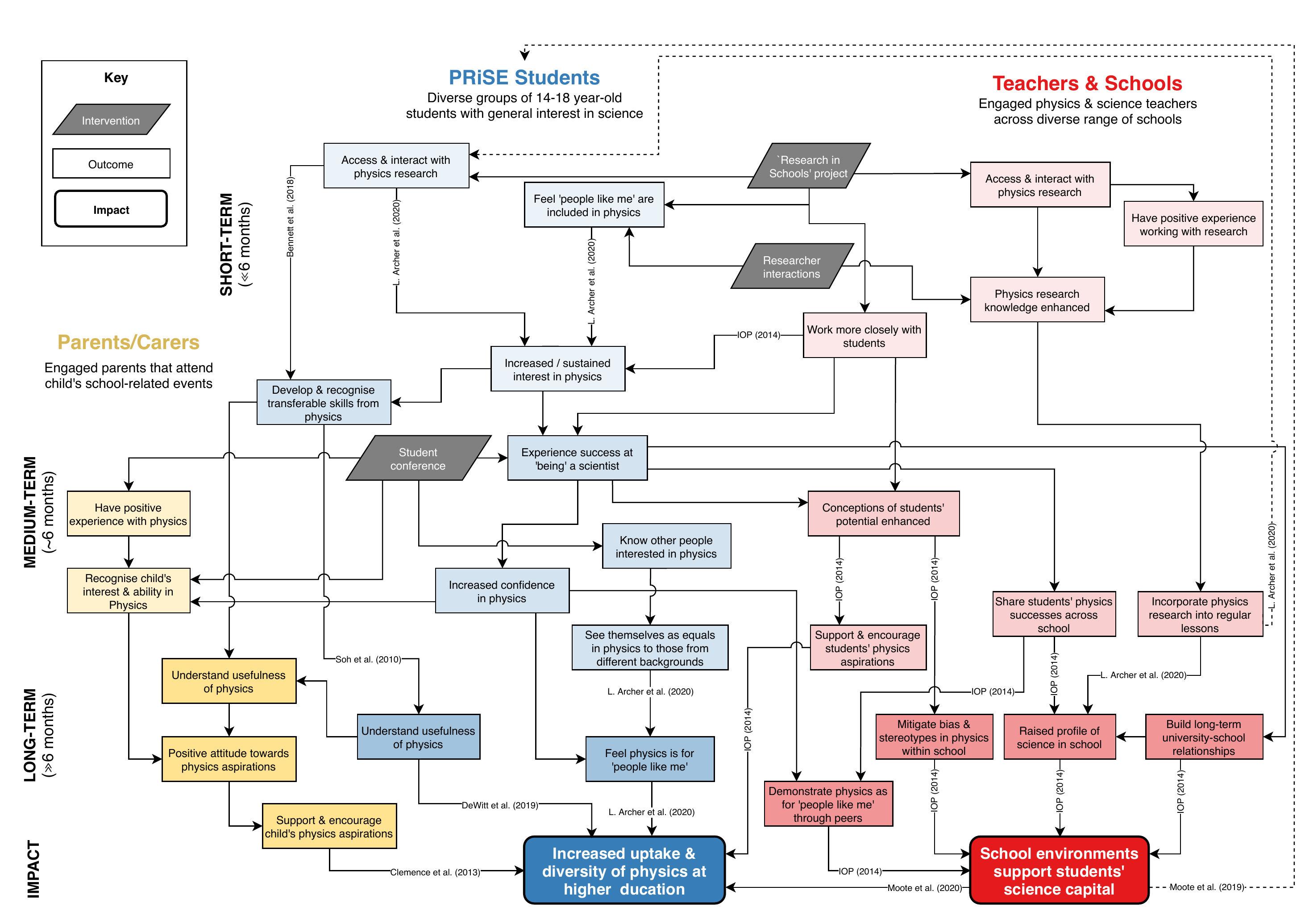}\caption{The Theory of Change for PRiSE concerning school students (blue),
their parents/carers (yellow), and their teachers and school environment
(both red). Shading indicates the timeframe of outcomes going from
lighter (short- and medium-term) to darker (long-term and impact)
with time running vertically from top to bottom. Arrows denote theorised
causal links between outcomes, accompanied by references where possible,
with dotted arrows indicating links that might feed back into successive
years of the programme.\label{fig:TOC}}
\end{sidewaysfigure}

Given this complexity, we contructed the aims of the PRiSE programme
through a theory of change (TOC; \citealp{sullivan06}). This approach
is recommended by several organisations and has been applied to other
STEM outreach/engagement programmes \citep[e.g.][]{davenport20}.
A TOC is designed to rationalise the intended outcomes and impacts
of an initiative by outlining causal links. The process of creating
a TOC works backwards, starting at the intended ultimate impact and
mapping the intermediate outcomes (both short-, medium-, and long-term)
that are thought to be required to enable that goal. The resulting
outcomes pathway (which may require iterating several times) should
be accompanied by the rationale for why specific connections exist
between different outcomes in the theory narrative along with any
underlying assumptions and potential barriers. We note that by no
means are all the causal links presented in a TOC guaranteed to occur,
however, by considering them during development programmes are more
likely to realise them. Figure~\ref{fig:TOC} displays the TOC for
PRiSE, which covers participating students (blue) as well as their
parents/carers (yellow) and their teachers and school environment
(both red).\nocite{archer20,bennett18,dewitt19,iop14,moote19,moote20,Soh2010,wellcome13}

The intended impact of PRiSE is to contribute towards the increased
uptake and diversity of physics at higher education. By serving students
near the end of their school-based educational journey, somewhat necessitated
by the content and style of open-ended `research in schools' projects,
the programme acts to support students' existing identity with science
in general and enhance, or at least maintain, physics aspirations
to help transform these into degree subject destinations --- a known
issue at this stage. Therefore, an assumption feeding into this TOC
is that PRiSE students will have an existing positive association
with science beforehand and we make no claim that the PRiSE approach
would be effective for students who are generally uninterested or
unengaged with STEM from the outset. Students' interest or enjoyment
in the subject as well as its perceived usefulness in a career are
key factors affecting degree choices \citep{dewitt19}, with students
(particularly girls) often thinking physics is less useful or relevant
\citep{murphy06}. Additionally, the stereotypes and school-based
practices associated with physics make many, even highly-able and
interested students, at this age conclude it is `not for me' \citep{archer20}.
PRiSE attempts to be a factor in addressing all of these factors in
some way. By interacting first-hand with ``real physics'' through
the projects and working with active researchers, students (especially
those from under-represented groups) should feel included and have
their interest in physics enhanced or at least sustained \citep{thorley20}.
By experiencing success at `being' a scientist and meeting similar
students from other schools, it is hoped their confidence will be
boosted leading to a feeling that physics is indeed for `people like
me' \citep{davenport20}. Furthermore, through working in new ways
students should develop numerous transferable skills \citep{bennett18}
which might help them recognise the usefulness of the subject \citep{Soh2010}.

Teachers are much stronger influences on students' aspirations than
university staff/students could ever be \citep{aspires13,aspires20}.
Experience from physics outreach officers (e.g. through discussions
via the South East Physics Network's Outreach and Public Engagement
and Ogden Trust's Outreach Officer programmes) has shown that most
teachers are more interested in activities for their students from
universities rather than continuing professional development opportunities,
which they may seek elsewhere. Therefore, opportunities for teachers'
development are integrated within the programme rather than being
a separate offering to schools. While the number of students working
on PRiSE may be relatively small, by influencing teachers through
our sustained programme, the aim is that the impacts of PRiSE can
be felt much more widely. Indeed, our hope is to affect the environments
within the diverse range of schools we work with on the programme
so that they are places that are able to support and nurture the science
capital \citep{aspires13,aspires20} of all their students, thereby
also contributing to our goal of increased uptake and diversity of
physics \citep{moote19,moote20}. \citet{iop14} recommends to help
achieve such an environment that schools should raise the overall
profile of science in school, endeavour to build long-term relationships
between pupils and role models, and ensure all teachers are aware
of the influence they can have on children\textquoteright s future
careers. We aim to further all these points by PRiSE providing more
collaborative working opportunities between teachers and students,
exciting success stories that teachers and students can share across
their schools, and the gateway to building longer-term relationships
between schools and the university. It is likely the teachers that
sign up for such a protracted programme and/or benefit the most from
it are relatively engaged generally. How to enhance the practice of
disengaged teachers is a challenge beyond the scope of PRiSE.

Finally, another major influence on young people's aspirations is
family, particularly parents or carers \citep[e.g.][]{wellcome13}.
Parental engagement is notoriously difficult within school-based programmes
in general \citep[see the review of][]{archer20r4a}, so we simply
aim to include parents/carers to celebrate in students' project work
at the end of the programme. These parents/carers are likely interested
in their children's education. It is hoped that by witnessing their
child's successes and development through physics, they will be more
positive, and thus supportive, towards physics aspirations going forward,
reinforcing the impacts of the programme.

This TOC for PRiSE was used in motivating the provision within the
framework. Evaluating the impact of PRiSE is beyond the scope of this
paper / supplementary material and is explored in a companion paper
\citep{archer_prise_impact20}. We stress that this TOC for PRiSE
does not exist in isolation and other TOCs which focus on different
stages of and aspects to a young person's learning ecology are required
to improve the overall issue of uptake and diversity in physics. For
example, \citet{davenport20} present a potential STEM outreach TOC
designed towards primary and lower secondary school students, with
outcomes focused more on STEM careers.

\section{Current projects\label{sec:Current-projects}}

Queen Mary University of London's (QMUL) physics research areas concern
astronomy (space science, planetary physics, and cosmology), particle
physics (the Standard Model and beyond via particle colliders and
neutrino observatories), condensed matter physics (e.g. material structure,
organic semiconductors, and applications thereof), and theoretical
physics (e.g. string theory, and scattering amplitudes). Of these,
it was decided to initially base PRiSE around the space and planetary
sciences as well as particle physics. These exciting topics are thought
to inspire awe in the public due to the \textquotedblleft big\textquotedblright{}
questions they address and the senses of scale and wonder beyond our
everyday experience \citep[cf.][]{madsen03,ippog}. However, exactly
how this science is conducted is not often well understood outside
of academia, particularly at school-level due to the lack of research
methods within current science teaching \citep[e.g.][]{hodson,braund07,yeoman17}.
Currently four projects have been developed for the PRiSE programme
at QMUL, which we briefly summarise here indicating key project personnel
referring to the roles mentioned in the main article (note the outreach
officer role for the entire programme was performed by the first author).
\begin{itemize}
\item \textbf{Scintillator Cosmic Ray Experiments into Atmospheric Muons
(SCREAM, 2014--2020) }was adapted by the outreach officer from a
dissertation project designed for undergraduates using a scintillator
-- photomultiplier tube muon detector \citep{coan,teachspin} fundamentally
similar to those found in current neutrino experiments such as SNO+,
where cosmic ray muons serve as an important background source that
can be used for calibration \citep{Alves2015}. Students calibrate
their borrowed detectors and collect counts of comic ray muons and
muon decays. Initially they use this data to perform a measurement
of muons' mean lifetime (using both software that comes with the detector
and programmes we have created especially) before progressing to a
wide variety of potential topics on these cosmic rays such as their
angular distributions or dependence on atmospheric/solar conditions.
Since detector usage and particle physics are part of most A-Level
physics syllabuses, it complements their studies. At present QMUL
only has four of these detectors as they are expensive (around \textsterling5,000),
which limits the number of schools we can work with each year. As
this project has been especially popular with teachers when signing
up, we have made it open only to schools that have successfully undertaken
a different project with us previously, also limiting how long they
can borrow a detector to a maximum of 4~years.\\
Key personnel: Dr Jeanne Wilson (project lead, 2014--2019), Prof
Peter Hobson (project lead, 2019--2020), Dr Martin Archer (researcher,
2015--2020)
\item \textbf{Magnetospheric Undulations Sonified Incorporating Citizen
Scientists (MUSICS, 2015--2020)} was created especially for PRiSE
by the project lead. Geostationary satellite data of the ``sounds
of space'', ultra-low frequency fluid plasma waves in Earth's magnetosphere,
have been made audible. Students are given this data on preloaded
USB flash drives and explore it through the act of listening (we also
provide them with earphones). Audacity audio software (\url{https://www.audacityteam.org}/)
is used to analyse any events identified, which can then be logged
in a specially created spreadsheet which performs some routine calculations.
We stress that students do not have to focus on the space plasma physics
aspects, which will largely be completely unfamiliar, but rather just
the waves topics that they cover in class both at GCSE and A-Level.
While students are given guidance on how to listen to and analyse
the waves, we do not prescribe to them exactly what to listen out
for as we are instead interested in what they pick out themselves.
This approach has already led to novel and unexpected scientific results
on the resonances present in Earth's magnetosphere during the recovery
phase of geomagnetic storms \citep{archer18}. Based on these results,
an optional `solar storms campaign' was created for 2019/20, providing
more concrete prescribed instructions to build up a dataset of similar
events followed by suggestions of unanswered questions about these
resonances that students could investigate. While a few schools followed
this route initially, they all eventually decided to go their own
way with it.\\
Key personnel: Dr Martin Archer (project lead and researcher, 2015--2020)
\item \textbf{Planet Hunting with Python (PHwP, 2016--2020) }was initially
developed by a post-viva PhD student, Dr Gavin Coleman, through a
one-day-per-week buyout over three months (funded by a grant obtained
by the outreach officer) and has subsequently been modified by the
project lead each year. The project aims to address the UK coding
agenda (the UK government\textquoteright s desire for more young people
to develop computer programming skills, \citealp[e.g.][]{dfe18})
by applying Python computer programming to data from NASA's Kepler
\citep{jenkins10} and later TESS \citep{ricker2015} missions, whereby
students write programmes to detect exoplanet transits. Transit photometry,
where an exoplanet blocks some of the star\textquoteright s light,
can be fairly easily understood by school students in terms of geometry
and the equations from A-Level physics. The students try to independently
implement each step laid out in their guide (period detection, phase
folding, model fitting, and parameter estimation) applied to specially
selected star systems chosen for their relative simplicity. Example
code is given to teachers. While extension activities are suggested,
so far very few students have progressed beyond the prescribed activities
within a single year, though some students have returned for a second
year making further progress.\\
Key personnel: Prof Richard Nelson (project lead, 2016--2020, and
researcher, 2017--2020), Dr Gavin Coleman (researcher, 2016--2017),
Dr Martin Archer (researcher, 2016--2017), Francesco Lovascio (researcher,
2020)
\item \textbf{ATLAS Open Data (2017--2020) }was adapted by the outreach
officer for PRiSE from a public resource produced by CERN aimed at
undergraduates \citep{atlas}. An undergraduate summer student, under
the instruction of the outreach officer, produced a guide so that
school students could build up to the documentation provided online
by CERN. At kick-off workshops students play a loaded dice game, developed
by the outreach officer and freely-available online as a resource
\citep{prise}, which serves as an analogy for why particle physicists
need to use statistical methods and big data in discovering new particles
such as the Higgs boson \citep{atlashiggs}. This leads into the main
activity, using CERN's interactive histograms to see how performing
cuts on the data increase/decrease the significance of the desired
signal, i.e. the Higgs, compared to the backgrounds. While the CERN
guides provide extensions by using their statistical software (ROOT)
for more detailed analysis, this has been beyond almost all PRiSE
students thus far, with most groups simply investigating the underlying
physics behind their chosen cuts to justify them.\\
Key personnel: Dr Eram Rizvi (project lead, 2017--2019), Dr Seth
Zenz (project lead, 2019--2020, and researcher, 2018--2020), Joe
Davies (researcher, 2019--2020)
\end{itemize}
Of the current projects at QMUL, only MUSICS at present has the scope
to lead to novel publishable scientific research, which it already
has done. The Kepler dataset has largely been mined of the clearest
exoplanets, often now requiring advanced machine learning techniques
for new discoveries \citep[e.g.][]{shallue18} which are currently
also being implemented on TESS. The other two projects have limitations
based on the equipment \citep{coan,teachspin} and amount of data
used (ATLAS Open Data's first release contained only a fifth of the
data used in the Higgs boson's discovery, \citealp{atlashiggs}, however,
more data was released in 2020). While this is not perhaps ideal,
it is due to the realities of pressures on academic staff time limiting
their ability to develop outreach projects from scratch \citep[e.g.][]{thorley16}.
\citet{archer_report17} recommended that the development and delivery
of `research in schools' projects should be distributed within each
research group sharing the load out amongst academics, post-docs,
and PhD students. This would allow more schools to participate without
overburdening individual researchers. However, this research group
buy-in has proven difficult to achieve at Queen Mary due to an overall
poor culture towards public engagement and outreach in their physics
department, a fairly common barrier to public engagement in general
\citep{burchell17}. Responsibilities have thus largely been falling
to only a few people per PRiSE project which has limited the number
of schools which could be involved each year. Nonetheless, there have
been some positive steps in the last year with project leads, along
with the outreach officer, being able to convince a few early career
researchers to help with delivery, which may indicate the department
slowly moving towards a more embedded approach. Institutions with
a more positive culture of public engagement and outreach likely could
support even more schools with research projects than has been possible
at QMUL.

\section{Implementation practicalities}

This section is designed to provide sufficient additional detail to
enable practitioners to fully understand how the PRiSE framework has
been implemented, with the aim of informing their schools engagement
practice.

\subsection{Activity stages}
\begin{itemize}
\item \textbf{Prescribed work:} This stage involves following a set of instructions
to undertake an experiment/activity designed to cover most aspects
of an investigation and to build their confidence in the project topic.
Students are still required to problem solve throughout these stages
and we purposely do not provide them with all the answers to prompt
this, though teachers are given guidance in their resources to support
student efforts. The stage is designed to enable students and teachers
to initially be able to access and interact with the research, enhance
teachers' knowledge, and hopefully make underrepresented groups feel
included in physics (cf. Figure~\ref{fig:TOC}).
\item \textbf{Independent project: }This continues in a similar way to the
prescribed work except now independently motivated. In visits and
webinars the question has been raised by students whether there is
a risk that they investigate the same thing as another group at a
different school, though given the broad scope of most of the PRiSE
projects so far this has rarely occurred. Potential research questions
are suggested in the guides provided, with further advice for teachers
on these being given in their versions, and students' ideas are discussed
during visits and/or webinars. During this stage we aim that students'
interest in physics is increased or sustained through pursuing their
own research questions and that they develop physics-related transferable
skills (cf. Figure~\ref{fig:TOC}). We encourage teachers to work
with their students throughout this stage, e.g. holding regular meetings,
which may help enhance their perceptions of students' ability (cf.
Figure~\ref{fig:TOC}).
\item \textbf{Writing up: }Near the end of the project students produce
either a scientific poster or talk to be presented at our annual conference.
Guidance on how to approach these is provided online as well as during
visits and webinars. We have found that with the researcher support
provided that all student groups that persist with project work to
this stage are able to produce a poster or talk, thereby experiencing
success at `being' a scientist (cf. Figure~\ref{fig:TOC}).
\end{itemize}

\subsection{Interventions}
\begin{itemize}
\item \textbf{Assignment:} Using existing teacher networks, such as the
Institute of Physics' Stimulating Physics Network \citep{hartley11}
and the Ogden Trust School Partnerships \citep{ogden} in the UK,
not only allows us access to schools from lower socio-economic areas
given the networks' focus but also act somewhat like a word-of-mouth
recommendation. We have found these networks to be more successful
at attracting new schools to the programme than our existing schools
events mailing lists. Teachers fill out an online form providing school
and contact details, project preferences, and the estimated number
of students who will be involved. Previously participating schools
have to reapply each year. Once applications are in we assess the
capacity of the programme (taking into account data about the schools)
and inform teachers before the summer break whether their school has
been allocated a project or not. Most schools are assigned only one
project, which both helps with logistics and makes it easier for teachers
to manage, and where possible we take into account their stated preferences
though this is not always possible given the researchers' workloads.
\item \textbf{Kick-off: }Due to constraints on time for academic members
of staff leading projects, the kick-offs are typically an evening
event on (university) campus. In some cases where schools could not
attend the event we have repeated it at their school. Projects led
by non-academic staff are usually hosted in-school, sometimes within
a normal lesson or at lunchtime/after-school depending on the teacher.
The events start with a 20--30~minute introductory talk by the project
lead concerning the underlying physics and research topic, leading
up to an overview of what the project is about, which is presented
to students as an opportunity that they can take advantage of if they
wish. The outreach officer then discusses the differences between
learning styles in the research project compared to their regular
classroom experience, how the project will work, the support available,
and how to go about obtaining this. The event ends with a hands-on
workshop for at least 20~minutes usually run by the outreach officer
and facilitated by researchers (though not always the project lead).
This workshop, which teachers as well as students are actively encouraged
to participate in, typically forms either the early part of or a lead
into the initial stage of the project work. Experts are on hand to
assist with any questions or initial troubles, with the aim of getting
the students and teachers to a place where they can continue this
work without too much extra help for the next month or so. Students
and teachers are given all the project's resources so they can begin/continue
their project work from this point on at their school. The outreach
officer will also have an informal chat with (particularly new) teachers
concerning how to go about undertaking and supervising the project,
answering any questions or concerns they may have with either the
science, activities, or project management.
\item \textbf{Visit: }These school visits by researchers are often administered
through a rolling Doodle (\url{http://www.doodle.com}) poll where
teachers can sign up to a session given the researcher's availability.
Schools taking advantage of this stage typically receive only one
visit, though if further demand is communicated we try to accommodate
one (or occasionally two) additional visits. The visits typically
last around an hour and occur around the stage where groups have finished
the prescribed activities and are thinking about or are in the early
stages of undertaking their independent research. They are very much
student-driven meetings, where the researcher asks the groups of students
to show what they have done, probes their understanding of this, puts
their work into the context of current research, provides answers
to any questions the students have, and gives advice on what the direction
and next steps with their specific project ideas might entail while
bearing in mind what methods/results may be achievable within the
timeframe of the project. Only active researchers have the necessary
expertise to draw upon in offering such bespoke, tailored guidance
to students and teachers working on projects in their research area.
With one project (ATLAS) and for a few schools on other projects it
has not been possible to have in-person visits for logistical reasons,
however, similar interactions were done via specifically arranged
Skype calls to the schools in these cases. Teachers are encouraged
to participate in these meetings and additionally further informal
chats (similar to those taking place during the kick-off) between
the researcher and teacher occur to help with their project supervision
and continuing professional development. These researcher interactions
with students and teachers are aimed at not only supporting project
work within the schools, but further enhancing teachers' knowledge
and students' sense of inclusion in physics (cf. Figure~\ref{fig:TOC}).
\item \textbf{Webinars:} One project (MUSICS) has experimented with additional
support to schools through monthly drop-in webinars between November--February,
providing further opportunities for students and teachers to ask questions
of the researcher and get advice on how to progress with their project
work in a similar manner to the visits. This was first trialled in
2018/19 through a Google Hangout simultaneously streamed on YouTube,
however, this option was later discontinued so a solution using a
Skype group call also broadcast to YouTube (via the NDI\textregistered\ 
feature and using Open Broadcast Software, \url{https://obsproject.com}/)
was implemented in 2019/20. The YouTube streams are unlisted so that
only project students with a link can access them, making the webinars
a safe space for them to discuss the project. While almost all students
and teachers preferred to simply join the YouTube stream and contribute
via its live chat facility, the rationale behind incorporating the
Hangout/Skype option was so that participants could directly talk
to the researcher and/or show their work. In 2018/19 the webinars
were organised in a somewhat ad hoc manner and due to technical limitations
the only way of communicating the links to join was via an email immediately
before the webinar. With the move to Skype we were able to create
a stable hyperlink to join the group as well as being able to embed
the YouTube events in advance on a password-protected webpage, both
of which allow for easier access to the webinars. In terms of organisation,
at the beginning of the 2019/20 academic year we sent out an online
form asking teachers to identify when might be the best times for
webinars. While the response rate for this was low (only four), we
used this to set a regular monthly schedule (in this case the first
Monday of each month at 4--5pm) which was communicated to teachers
far in advance. All these changes considerably increased the uptake
of webinars: 10 out of 14 schools participated in at least one webinar
in 2019/20 compared to only 2 out of 14 the previous year. The rationale
behind webinars is that they further support the projects and their
aims in a way that makes efficient use of researchers' time.
\item \textbf{Ad hoc:} We explain at the kick-off meetings that when students
get stuck at any point (which they invariably will do due to the nature
of research) they should try to first tackle this themself, before
discussing in their groups, and then raising with their teacher. In
general, teachers act as the primary contact to students offering
encouragement and any support or advice they can. If students' questions
go beyond what their teacher can answer and is not covered by the
teacher guides we provide, the teachers should get in touch through
the outreach officer. This is done not only for logistical and safeguarding
reasons, but also provides further opportunities for university--teacher
dialogues that can contribute to their professional development. Some
teachers, however, instruct their students to email directly. Not
all schools require this option of further support and we have not
yet been overloaded with additional questions. Only in a few cases
has the outreach officer not been able to directly answer the question,
subsequently passing it on to a relevant researcher to answer, though
in general this would depend on the background and experience of the
outreach officer.
\item \textbf{Comments:} These are currently given by the outreach officer,
though in general this would depend on their background/experience
and could instead be done by the relevant researcher role(s). Teachers
(or students directly) email their work to the outreach officer and
receive annotated versions back the week before the final deadline,
allowing the students at least a few days to implement any changes.
\item \textbf{Conference:} The evening is primarily based around oral and
poster presentations by the students. Food and drink are provided
during the poster session and we also put on various physics demonstrations.
At the end of the evening all student groups are congratulated and
given a thank you letter, with a select number of groups highlighted
by researcher judges also receiving prizes in the form of various
science gadgets (some prize winners have also had the opportunity
to present their research at a national student conference hosted
by the Royal Society). As of 2019 we limited the number of talk slots
available to four in total, both for time and so each topic can be
covered. Schools are only able to solicit one talk (where desired)
by providing a title and abstract in advance (early March), with the
decisions of who will present being made that same week. There are
currently no limits on the number of posters a school can enter into
the conference. The event further enables students to experience success
at `being' a scientist as well as getting to know other people, outside
of their school from a variety of backgrounds, with interests in physics.
It is hoped that this might lead to increased confidence, seeing themselves
as equals in physics, and ultimately that physics is something for
`people like me' (cf. Figure~\ref{fig:TOC}). The conference provides
parents/carers in attendance the opportunity to have a positive experience
with physics and witness their child's interest and ability in the
subject, which could in the long-term result in their supporting and
encouraging their child's physics aspirations (cf. Figure~\ref{fig:TOC}).
Finally, from teachers' perspectives the conference lends an avenue
for them to support and encourage their students as well as share
successes across their school (cf. Figure~\ref{fig:TOC}).
\end{itemize}
The outreach officer typically sends updates and reminders about these
possible interventions to all teachers involved fortnightly throughout
the programme. This frequency attempts to tackle teachers' generally
low levels of response to a single email (\citealp{sousasilva17},
also reported teachers' generally poor communication through ORBYTS).
In addition to email communications, we also set up a password-protected
teacher area on our website in 2019 which always contains the latest
information on the programme. It appears from website analytics that
teachers have used this to some extent (there were 76 unique page
views amongst the 38~teachers involved that year), though we are
unsure which teachers these were and how often they visited the page
throughout the programme. For any on campus events, schools travel
to Queen Mary using London's extensive public transport network and
we are unable to offer schools compensation for travel expenses due
to limited funding. Outside of London or other well-connected cities
this travel may be a greater barrier to participation than in our
case so may need further consideration.

\subsection{Resources}
\begin{itemize}
\item \textbf{Project poster/flyer:} At the start of the academic year we
send posters/flyers to teachers to help attract attention to the project
within their school. Some teachers had been making their own such
adverts previously, which motivated us to create these.
\item \textbf{Project guide: }Our guides are presented in the style of an
academic paper. Printed copies of these are given out at the kick-off
event and electronic version can also be found on the project's page
on our website. These serve as an introduction, providing enough information
for students to start working on their project and be something they
can refer back to throughout. However, the guide is not intended to
be exhaustive (that would be impossible given the open-ended nature
of the projects) and students are made aware that we expect them to
read additional materials as they progress. Throughout the guides
there are exercises and discussion questions for students to consider,
designed to help them think more critically about their project work.
Teachers' guides include answers to the exercises, hints and tips
about different methods, common pitfalls that students make etc. and
these are distributed to teachers at the kick-off, via email, and
also stored on the password-protected teacher webpage. These project
guides have been updated every single year based on feedback and professional
experience, which is straightforward to do given the chosen style
compared to say a more illustrated glossy guide that would require
a professional designer each time.
\item \textbf{Project webpage: }Including previous work on these had been
suggested by students and teachers in feedback for a few years as
something that would be helpful. Some projects have also produced
videos which provide further information on the science / research
area or demonstrate how to use tools provided for the project, which
have been included on these pages.
\item \textbf{`How to' guide:} These articles have been produced by the
outreach officer. The most used of these are the guides on producing
scientific talks and posters that we point students and teachers to
ahead of the conference. While articles in this section designed for
teachers have also been planned, which would highlight elements of
good practice that have emerged from other teachers on how to successfully
integrate and nurture project work within their schools outside of
the support offered by researchers, we have not had sufficient time
or detailed input from teachers to be able to co-create these yet.
\end{itemize}
Examples of all these resources are given in section~\ref{sec:Example-resources}.

\subsection{Funding}

The operational costs of PRiSE's interventions have largely been covered
by QMUL's departmental physics outreach budget. Running the kick-off
events and conferences is comparable in cost ($\sim$\textsterling3,000
p.a. at PRiSE's current scale) to many summer school programmes aimed
at high school students which universities offer and often absorb
the cost of. However, as noted in the review of \citet{archer20r4a},
summer schools have severely restricted places ($\sim$10--30 students)
and limitations on impacts upon students. The PRiSE model thus potentially
offers much greater value for money.

The programme management falls within the scope of the funded outreach
officer post, a role which many university departments employ \citep[e.g.][]{ogden}.
Whether to pay early career researchers or assign workload allocations
to academics in delivering projects would realistically be down to
the policy/strategy of the department or institution. In the case
of Queen Mary, we have opted to only offer pay to PhD student researchers
(sourced from the department's outreach budget) to try to increase
uptake of engagement. Workload allocation in outreach / public engagement
for academics is dedicated only to roles aimed at embedding engagement
throughout the department, e.g. champions in research groups who help
recruit their colleagues to contribute to PRiSE. Delivery of engagement
is considered an expectation of the role of an academic, though significant
contributions may be used as criteria for promotions. Grants have
been obtained to assist with project development, as detailed in section~\ref{sec:Current-projects}.

If institutional funds are not available for the delivery of a `research
in schools' programme, we would recommend either costing these into
research grants as impact-related costs or employing the ORBYTS funding
model of charging independent schools to allow both them and a few
less-resourced schools to participate \citep{sousasilva17}.

\section{Example resources\label{sec:Example-resources}}

Finally we give examples of the resources within the PRiSE framework.
We display the posters (A3 size) provided to teachers to advertise
projects to their students in Figure~\ref{fig:posters}. Leaflets
(A5) size used similar designs and content. We also show a project
webpage (Figure~\ref{fig:project-webpage}) and teacher guide (Figure~\ref{fig:guide}),
in both case for the MUSICS project. Student guides are identical
but do not include the red text, which is for teachers only. Finally,
a `how to' style online guide for students is given in Figure~\ref{fig:how-to},
specifically the one on creating research posters.
\begin{center}
\includegraphics[scale=0.65]{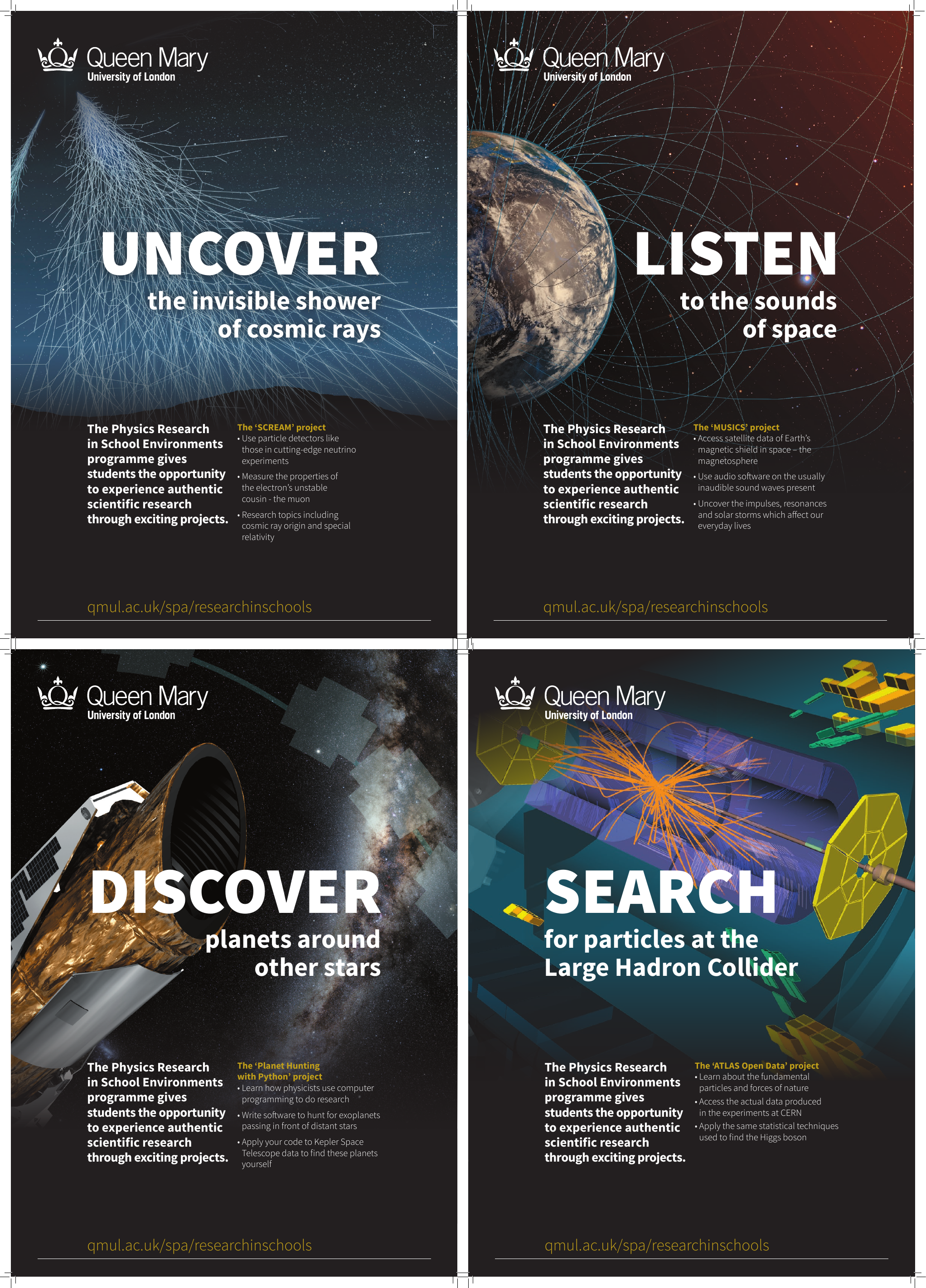}\captionof{figure}{Posters
for advertising the four PRiSE projects at QMUL.\label{fig:posters}}
\par\end{center}

\begin{center}
{\fboxsep 0pt\fbox{\includegraphics[scale=0.75]{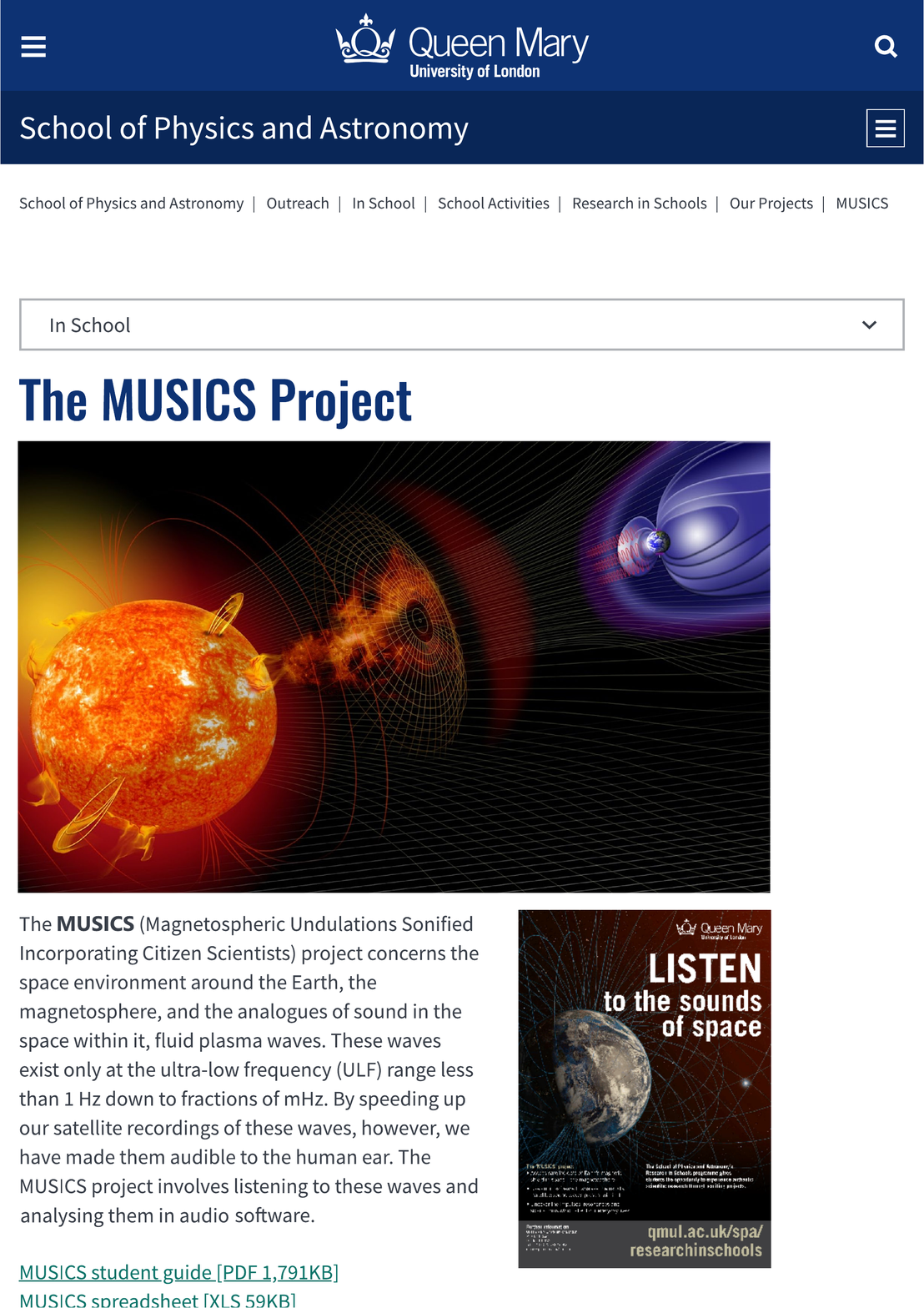}}}\captionof{figure}{Example
PRiSE project webpage\label{fig:project-webpage}.}\includepdf[pages=2-3,frame, scale=0.75,pagecommand={}]{project_webpage.pdf}
\par\end{center}

\begin{center}
{\fboxsep 0pt\fbox{\includegraphics[scale=0.75]{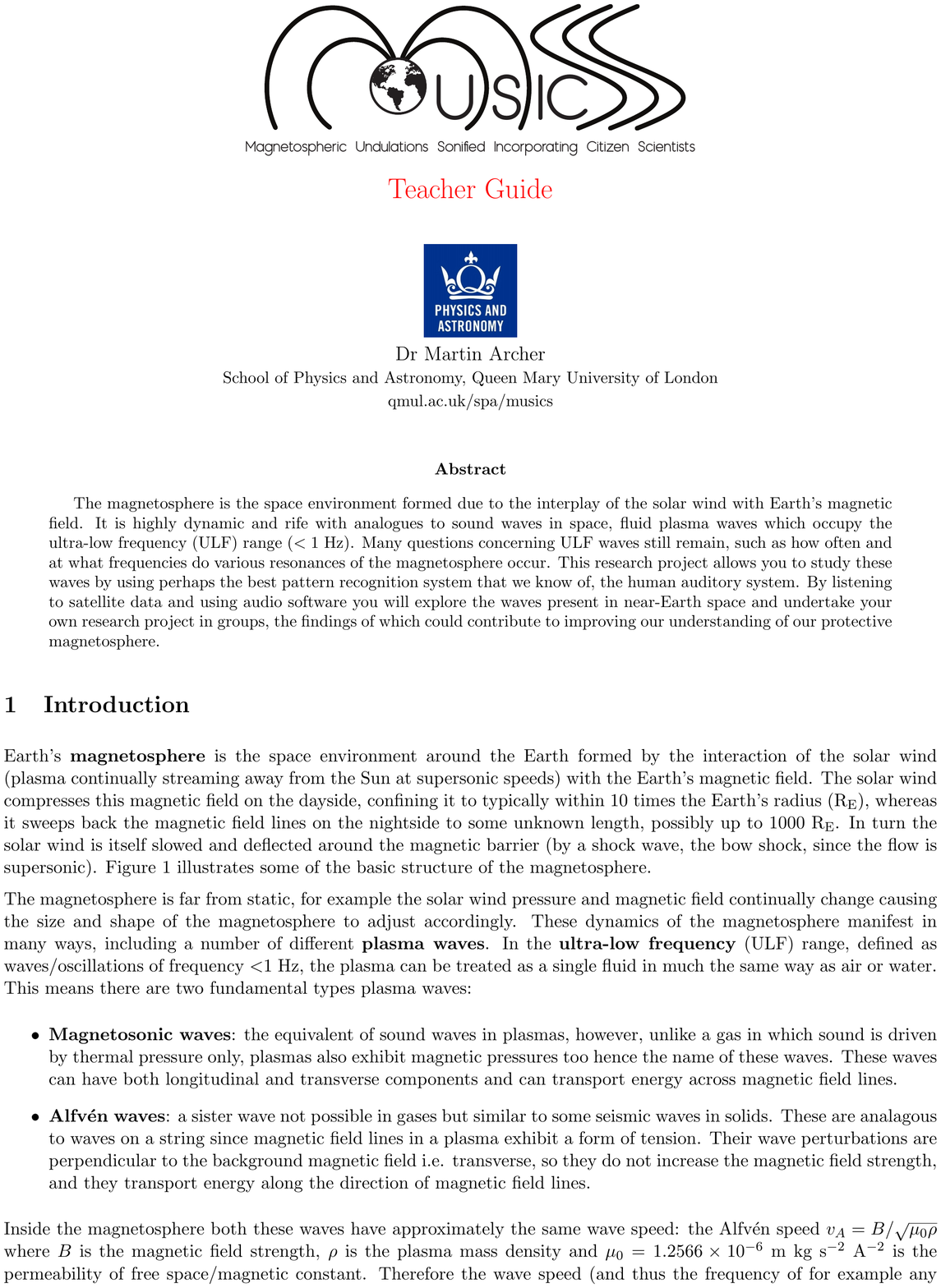}}}\captionof{figure}{Example
PRiSE project teacher guide\label{fig:guide}.}\includepdf[pages=2-10,frame, scale=0.75,pagecommand={}]{ULF_project_v6_teacher_supplemen.pdf}
\par\end{center}

\begin{center}
{\fboxsep 0pt\fbox{\includegraphics[scale=0.75]{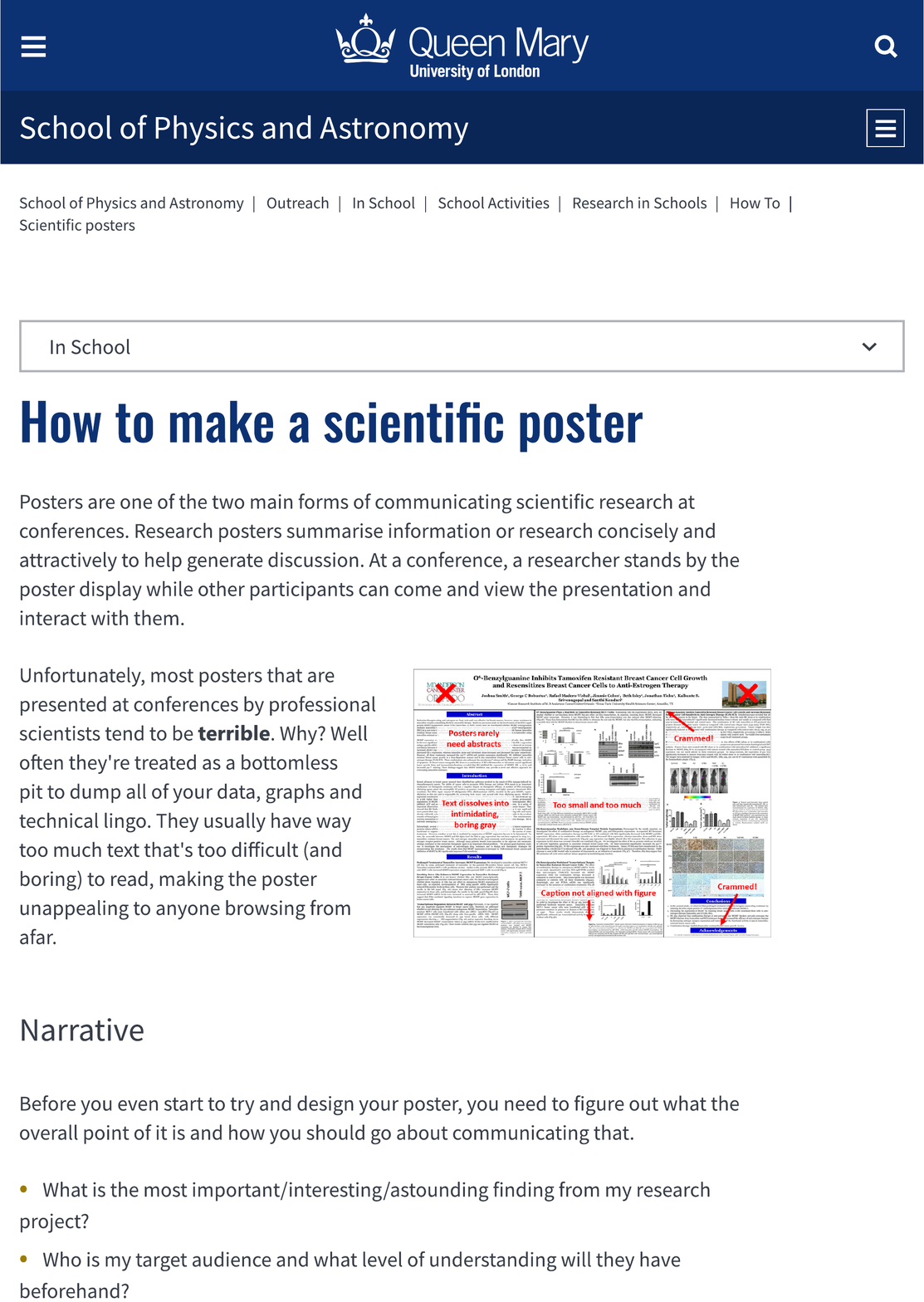}}}\captionof{figure}{Example
`how to' guide\label{fig:how-to}.}\includepdf[pages=2-4,frame, scale=0.75,pagecommand={}]{how_to.pdf}
\par\end{center}

\DeclareRobustCommand{\disambiguate}[3]{#1}

\bibliographystyle{copernicus}
\bibliography{prise2020}